\documentclass[twocolumn]{aastex62}
\usepackage{CJK}
\usepackage{natbib}
\usepackage{graphicx}
\usepackage {amsmath,amssymb,eqnarray}
\usepackage{enumerate}
\usepackage{enumitem}
\usepackage{mathtools}

\newcommand{\hi}{{\rm H}{\textsc i}}
\newcommand{\angstrom}{\text{\normalfont\AA}}

\begin{document}

\title{ Suppressed or enhanced central star formation rates in late-type barred galaxies}

\author[0000-0002-6593-8820]{Jing Wang}
\affiliation{ Kavli Institute for Astronomy and Astrophysics, Peking University, Beijing 100871, China}

\author[0000-0001-6079-1332]{E. Athanassoula}
\affiliation{Aix Marseille Universit{\' e}, CNRS, LAM, Laboratoire d'Astrophysique de Marseille, Marseille, France}

\author[0000-0002-3462-4175]{ Si-Yue Yu}
\affiliation{ Kavli Institute for Astronomy and Astrophysics, Peking University, Beijing 100871, China}
\affiliation{ Department of Astronomy, School of Physics, Peking University, Beijing 100871, People's Republic of China}

\author[0000-0002-4569-016X]{Christian Wolf}
\affiliation{ Research School of Astronomy and Astrophysics, Australian National University, Canberra ACT 2611, Australia }

\author[0000-0003-2015-777X]{Li Shao}
\affiliation{National Astronomical Observatories, Chinese Academy of Sciences, 20A Datun Road, Chaoyang District, Beijing, China}

\author[0000-0003-1015-5367]{Hua Gao}
\affiliation{ Kavli Institute for Astronomy and Astrophysics, Peking University, Beijing 100871, China}
\affiliation{ Department of Astronomy, School of Physics, Peking University, Beijing 100871, People's Republic of China}

\author[0000-0002-8791-2138]{T. H. Randriamampandry}
\affiliation{ Kavli Institute for Astronomy and Astrophysics, Peking University, Beijing 100871, China}

\begin{abstract}
Bars in disc-dominated galaxies are able to drive gas inflow inside the corotation radius, thus enhancing the central star formation rate (SFR). Previous work, however, has found that disc-dominated galaxies with centrally suppressed SFR frequently host a bar. Here we investigate possible causes for the suppression of central SFR in such cases. We compare physical properties of a sample of disc-dominated barred galaxies with high central SFR (HC galaxies) with those of a sample of disc-dominated barred galaxies with low central SFR (LC galaxies). We find that the two samples have on average similar \hi{} content and bars of similar strength. But we also find that the HCs have bluer colors than LCs, and that outside the bar region they host stronger spiral arms than the LCs where closed rings are more often seen. We discuss and evaluate the possible causes for the suppression of the central SFR in the LC galaxies as opposed to its enhancement in the HC galaxies.
\end{abstract}


\section{Introduction}
\label{sec:introduction}

Disc galaxies evolve both in their morphology and in their stellar population. In the nearby universe, the average evolution of galaxies is no longer violent or driven by major mergers as at higher redshift \citep[e.g.][]{Kraljic12}. The ways in which nearby galaxies evolve are predominantly secular, via disc instabilities. Bars are believed to play a significant role in the secular evolution of disc galaxies (for reviews of the theoretical and observational aspects see \citealt{Athanassoula13a} and \citealt{Kormendy13}, and references therein).

One of the most important mechanisms related to bars is that the non-axisymmetric potential of strong bars exerts torques on the gas, and drives it toward the center of galaxies (\citealt{Schwarz81}; \citealt{Athanassoula92b}, hereafter A92b; \citealp{Regan99, Sormani15, Fragkoudi16}). This process builds a concentration of cold gas at the center \citep[e.g.][]{Sakamoto99, Sheth05}, and triggers vigorous star formation therein \citep{Ellison11,CatalanTorrecilla17}. The existence of strong bars may account for nearly half of the central starbursts in massive galaxies at low redshift \citep[][W12 hereafter]{Wang12}. As the central concentration of stellar mass is built up via star formation, the morphology of the galaxy also changes \citep[but see][for barred early-type disk galaxies in the local universe]{Laurikainen07}. Hence, bars play an effective role in shaping the age and morphology of disk galaxies. 

Several studies came to the conclusion that bars are able to survive the dissolution effects induced by typical central mass concentrations, i.e., they are long-lived disc phenomena \citep[e.g.,][and references therein]{Shen04, Athanassoula05, Debattista06, Berentzen07, Aguerri09, Kraljic12, Athanassoula13}, so they would play a major role in the secular evolution of their host galaxies. However, others reached the opposite conclusion \citep{Bournaud02, Bournaud05}. Note, however, that central mass concentrations can weaken the bar, even if they can not completely destroy it \citep{Athanassoula13a}. Furthermore, major or extensive minor mergers can effectively destroy the bar \citep{Pfenniger91, Athanassoula99, Berentzen03, Sheth12}. Which of these has the upper hand can be found by following the bar fraction of early-type disc galaxies as a function of time. \citet{Sheth08} and \citet{Melvin14}  found that this fraction remains constant or increases with decreasing redshift, or look back time, which argues in favour of long-lived bars. Note that bars may play an even more important role in the future evolution of late-type disc galaxies, as the fraction of them that host bars continuously increases from $z\sim 0.84$ to the present day \citep{Sheth08}. 

Galaxies that host strong bars are not always found to have high central concentrations of cold gas or star formation rate \citep[SFR,][]{Martinet97, Sheth05, Cullen07, Fisher13}. Around one third of barred galaxies are found to have very low densities of cold gas at the center, regardless of their Hubble types \citep{Sheth05}. There are several possible reasons for the lack of cold gas in the center of barred galaxies. First of all, in order for bars to push gas inward, the gas needs to be available within the radial range where bars are effective \citep{Kuno07}. In barred disc galaxies, gas is not only driven inward but also outward, with a division line close to the corotation \citep[CR;][]{Kalnajs78, Bournaud02, Combes08}. Thus the flow of gas strongly depends on the radial distribution of the gas. Secondly, abundant gas might be present at the center before, but has been efficiently consumed by a past starburst. If we have a large observational sample, we may witness the smoking gun of this process by selecting the post-starburst galaxies. The central star formation may revive if new gas comes in. Some models predict that the onset and quenching of star formation at the center of bars occurs periodically, regulated by a balance between the inflow rate and the central concentration of mass \citep{Krumholz15}. Finally, the complex orbits of the gas strongly depend on a variety of parameters and can rarely reach the center under certain circumstances (\citealt{Athanassoula92}; A92b; \citealp{Quillen95, Sheth02}). For example, the inner Lindblad Resonance (hereafter ILR) may stop the gas before it reaches the centre if there is no nuclear non-axisymmetric structures to further tunnel the gas inward. 

Bars themselves also evolve while they drive the secular evolution of their host galaxies. Simulations suggest that bars transport angular momentum to the outer discs, the classical bulges and dark matter halos, so that they slow down and grow stronger and longer \citep{Athanassoula02a, Athanassoula03}; gas transfers part of its angular momentum to the stellar bar and weakens it \citep{Bournaud02, Berentzen07, Athanassoula13}. Thus, at a fixed central concentration of the optical light, the length of bars over the disc size is found to be correlated with the color of galaxies (W12). The strength of bars is also related to other morphological features: strong bars are shallower in the surface brightness radial distributions, more rectangular in shape and the contrasts from the underlying discs are stronger than weak bars \citep{Kim15, Kim16}. In extreme cases, the majority of the stars initially in the disc may be trapped to the bar \citep{Gadotti03, Gao17}. Stronger bars drive stronger gas inflows \citep{Sheth05, Kim12}, in agreement with simulation results \citep{Athanassoula94}.
 
W12 coherently studied the enhancement and suppression of star formation in barred galaxies within one sample, which is selected by the stellar mass, optical central concentration, and redshift. They found that at the same concentration, disc galaxies which have suppressed or enhanced central SFR tend to host more bars than galaxies which have the intermediate central SFR levels. This result seems to imply a picture where the bars play a role in the star formation quenching at the center of galaxies. However, as W12 focused on the built-up of the central mass concentration of galaxies, the properties of the barred galaxies with centrally suppressed star formation were not thoroughly investigated. It remains unclear whether the suppression of star formation is limited to the centre, whether the suppression is temporary, whether the total amount of cold gas is significantly reduced in these centrally suppressed galaxies, whether the bars in the two types of galaxies are of the same type, and whether another more intrinsic parameter should be responsible for both the quenching of star formation and the formation of bars. Finally, could the enhancement in bar fraction a consequence instead of a cause of the suppressed central star formation? The answers to these questions may be directly related to the dynamical evolution of bars and stellar evolution of galaxies. We aim to provide better understanding of these questions in this paper. 

The paper is organized as follows. We build samples of barred galaxies that have enhanced and suppressed central star formation, and describe the data to be analyzed in Sect.~\ref{sec:sample}. We compare properties of these different samples in Sect.~\ref{sec:result}. We use fiber spectral indices to infer the central star formation history, color profiles to characterize the radial range of SFR suppression or enhancement,  \hi-richness to indicate the availability of gas reservoirs, and different radial profiles to indicate the strength of the bars and spiral arms. We discuss the results in Sect.~\ref{sec:discussion}. Throughout this paper, we assume a Chabrier initial mass function \citep{Chabrier03}, and a $\Lambda$CDM cosmology with $\Omega_{m}=0.3$, $\Omega_{\lambda}=0.7$ and $h=0.7$. We will refer to as ``central region'' the region where the central SFR is clearly enhanced or suppressed. The region beyond it, but within the bar radius will be referred to as the main bar region, and the region outside that will be the ``outer region''.

\section{Sample and analysis}
\label{sec:sample}
\subsection{Parameters and the parent sample}
\label{sec:parent_sample}
We select galaxies based on catalogs from the seventh data release (DR7) of the Sloan Digital Sky Survey \citep[SDSS,][]{York00}, and the sixth data release (GR6) of Galaxy Evolution Explorer \citep[GALEX,][]{Martin05}. We extract parameters from the SDSS MPA-JHU catalog\footnote{http://www.mpa-garching.mpg.de/SDSS/DR7/}, including redshift ($z$), stellar mass ($M_*$), axis ratio ($b/a$), and semi-major axis length ($R_{25}$) of the 25\,mag~arcsec$^{-2}$ isophotal ellipse, and radius that encloses 90\% ($R_{90}$) and 50\% ($R_{50}$) of the total flux. We also take the indices 4000-$\angstrom$ break ($D_n(4000)$) and  Balmer absorption index ($H\delta_A$), the central star formation rate (SFR) and central stellar mass within the SDSS spectroscopic fiber region (3\,\arcsec{} in diameter, $\sim$1\,kpc in radius). The central SFR was derived from attenuation corrected H$\alpha$ luminosities for star-forming  galaxies, and from a $D_n(4000)$ based formula for non-star-forming (passive, composite or active galactic nuclei (AGN)-hosting) galaxies \citep{Brinchmann04}. 

We calculate the central stellar surface densities ($\Sigma_{*,ct}$) with the central stellar mass.
The effective stellar mass surface densities, $\mu_*$ is calculated as $0.5M_*/(\pi R_{50,i}^2)$, where $i$ refers to the $i$-band. The axis ratio is calculated as $b/a$ based on the $g$-band measurements.  The concentration index $R_{90}/R_{50}$ is based on the $r$-band measurements. 

The total SFR of each galaxy has been derived in W12, by fitting the spectral energy distribution (SED) of the optical (from SDSS) and ultraviolet (from GALEX) bands with a library of model SEDs which were built with the stellar population synthesis method and based on the spectral templates from \citet{BC03}. A range of star formation histories (SFH), metallicities and dust attenuations were assumed to ensure physically meaningful solutions. More details can be found in W12, \citet{Wang11}, and \citet{Saintonge11}. The SFR central concentration is defined as $C(SF)=\log (({\rm SFR}/M_*)_{\rm center}/({\rm SFR}/M_*)_{\rm total})$, where ``center'' refers to measurements within the SDSS fiber, and ``total'' refers to the global measurements. The $C({\rm SF})$ enhancement, $\Delta C({\rm SF})$ is calculated as the excess of $C({\rm SF})$ over the average of galaxies that have similar $M_*$ (differences $<0.2$ dex), $\Sigma_{*,ct}$ (differences $<0.2$ dex), $R_{90}/R_{50}$ (differences $<0.15$), and redshift (differences $<0.005$). So that a galaxy with a negative $\Delta C({\rm SF})$ has a suppressed central SFR compared to galaxies with similar stellar surface density distributions. The different indicators used to calculate the central and total SFR may cause systematic uncertainties in $\Delta C({\rm SF})$, but we will show in Sec.~\ref{sec:result_global} and ~\ref{sec:result_colorprof} that sub-samples selected with different ranges of $\Delta C({\rm SF})$ show distinct central $D_n(4000)$, $H\delta_A$, and $g-r$ color slopes, but less different global SFR. Hence on average $\Delta C({\rm SF})$ works well in selecting samples with different SFR concentrations. Large samples of SFR radial profiles estimated from uniform indicators could ultimately minimize this type of uncertainties, and should be available when the MaNGA data is fully released in the future.

After excluding all the galaxies with significantly asymmetric morphologies ($\sim3.4\%$), we select all the galaxies with $0.02<z<0.05$,  $M_*>10^{10}\,M_{\odot}$, $\log {\rm SFR}/M_*>-11$, $b/a>0.75$, $R_{90}/R_{50}<2.2$. This makes the {\it parent sample of 1478 massive, face-on, star-forming, and disc-dominated galaxies}. 

The maximum $R_{90}/R_{50}\approx 2.2$ of the main sample is much lower than the typical threshold of 2.6 for selecting disc-dominated galaxies (see SDSS-based studies, e.g. \citealt{Kauffmann03}). By selecting galaxies with very low $R_{90}/R_{50}$, we minimize potential influence of bulges. From \citep{Nair10}, most ($>90\%$) of the galaxies with $R_{90}/R_{50}<2.2$ have a Hubble type later than S(B)b. But whenever possible in this paper, we avoid discussions based on the relatively subjectively defined Hubble-sequence, so that the results could be more conveniently extended to other samples at low and high redshifts, and be compared to simulations in the future. 

We list the names of parameters and the abbreviations of terms which are used in this paper in the Appendix.

\subsection{The barred sample}
\label{sec:bar_sample}
The bar structures have been identified in W12, based on the standard method of analyzing the shapes of ellipticity ($e=1-b/a$, where $b$ and $a$ are the semi-minor and semi-major axis of an isophotal ellipse) and position angle radial profiles. The method is based on the fact that, in a galaxy which hosts a strong bar, the ellipticity in general rises continuously as a function of radius within the bar, reaches its maximum ($e_{\rm bar}$) at a point which we will associate with the end of the bar ($r_{\rm bar}$, the semi-major axis), and then drops to reach the ellipticity of the disc region; the position angle remains roughly constant as a function of radius within the bar, and suddenly changes at the end of the bar. W12 showed that, the strong bars (with $e_{\rm bar}>0.5$ and $r_{\rm bar}>2.5$ kpc) can be more reliably identified (with a reliabilities of $>70\%$) than the weak bars (with $e_{\rm bar}<0.5$ or $r_{\rm bar}<2.5$ kpc), and only strong bars have significant effect in enhancing the central star formation activities. Following W12, we will only discuss the galaxies with strong bars, and refer to them (388 galaxies) as the ``barred sample'' hereafter in this paper. 

\begin{figure*}
\includegraphics[width=17cm]{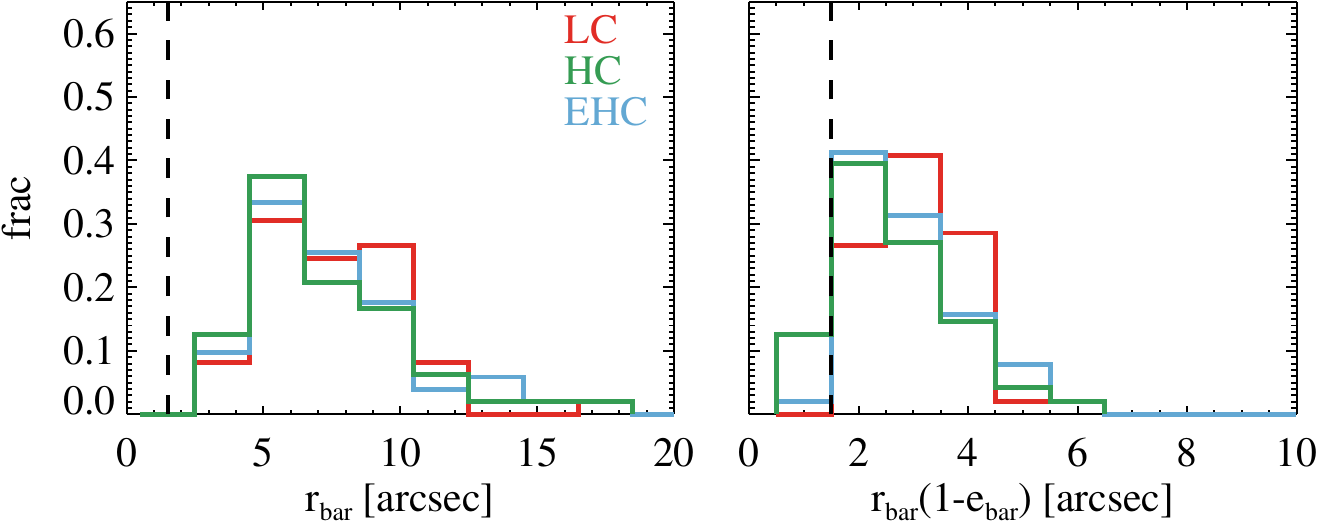}
\caption{The distribution of semi-major and semi-minor axes of bars in the three barred samples. The left and right panels plot the distribution of semi-major axes ($r_{\rm bar}$) and semi-minor axes ($r_{\rm bar}(1-e_{\rm bar})$) of bars, respectively.
The red, green, and blue colors are for the LC, HC, and EHC samples (see Sect.~\ref{sec:subsample} for details), respectively. The dashed line mark the SDSS fiber radius of 1.5 arcsec.}
\label{fig:barhist}
\end{figure*}

From Fig.~\ref{fig:barhist}, we can see that the SDSS fiber size is smaller than the semi-major and semi-minor axes of bars in most of the galaxies analyzed in this paper.

\subsection{The LC, HC, and EHC samples }
\label{sec:subsample}
The aim of this paper is to understand possible causes for the suppression of central SFR in a sample of strongly barred disc-dominated galaxies. 
To demonstrate the motivation of criteria used for selecting sample, we show the relation between $f_{\rm bar}$ and $\Delta C({\rm SF})$ (see also W12). 

\begin{figure} 
\includegraphics[width=8.cm]{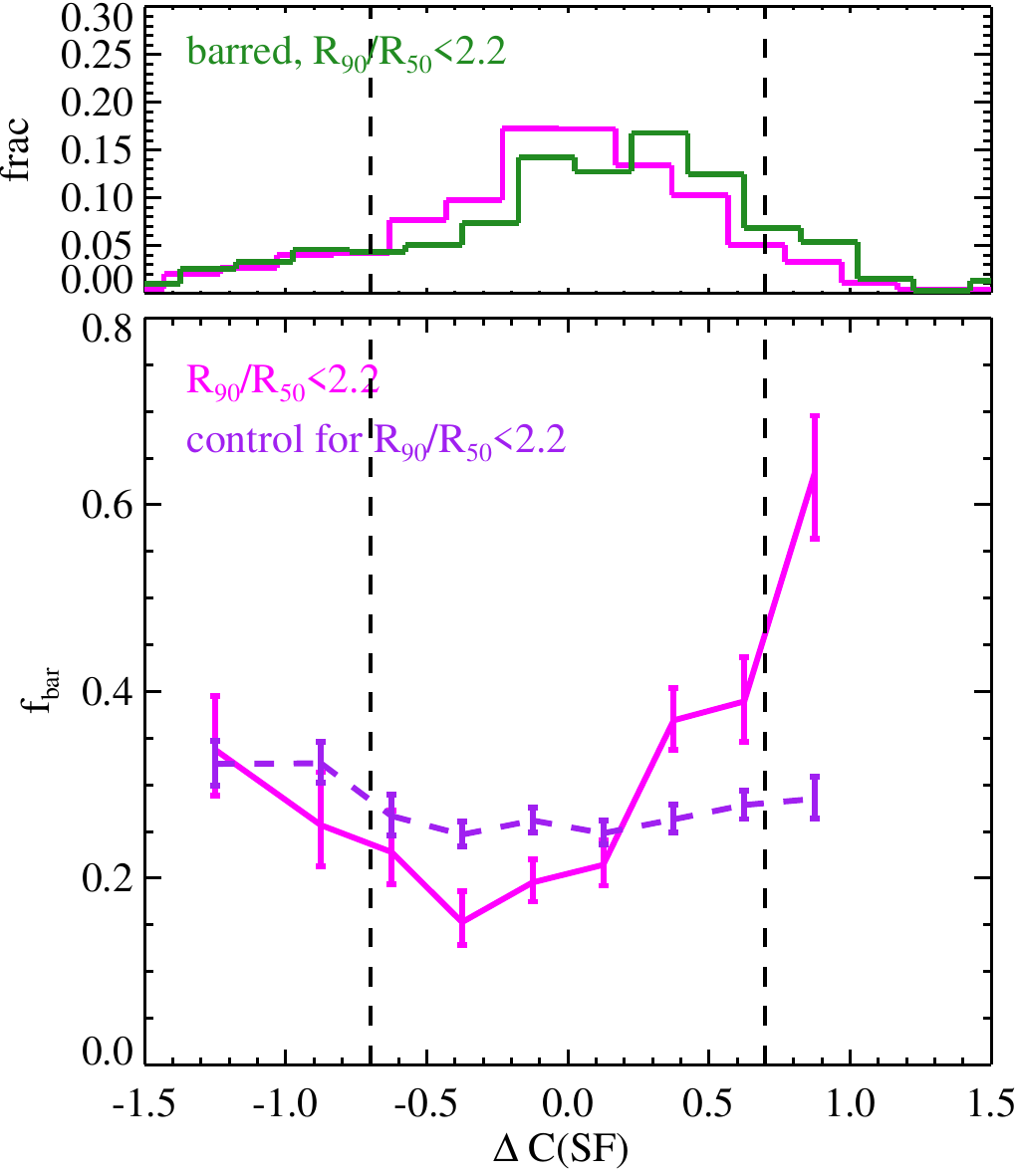}
\caption{{\it Top}: the distribution of relative central SFR concentration ($\Delta C({\rm SF})$, see Sec.~\ref{sec:parent_sample}) for all galaxies with $R_{90}/R_{50}<2.2$ (magenta) and for barred galaxies with $R_{90}/R_{50}<2.2$ (green). {\it Bottom}: the distribution of frequency of galaxies hosting bars ($f_{\rm bar}$) as a function of $\Delta C({\rm SF})$. 
In the bottom panel, the parent galaxies are plotted in magenta, and their control samples (matched in $M_*$ and $R_{90}/R_{50}$) are plotted in purple. These two samples are only used in this figure to support our selection of analysis samples LC, HC, and EHC in Sec.~\ref{sec:subsample}. The parent sample (magenta) shows a rise of $f_{\rm bar}$ toward low $\Delta C({\rm SF})$ (to the left of the first dashed line), which cannot be fully explained by a dependence of these two parameters on $M_*$ or $R_{90}/R_{50}$, for the control sample (purple) shows a much more flattened trend. The error bars represent the 0.683 (1-$\sigma$) binomial confidence intervals \citep{Cameron11}.} The two dashed vertical lines mark the selection of the LC and the EHC galaxies (see Sec.~\ref{sec:subsample}). 
\label{fig:fbar_cssfr}
\end{figure}

In Fig.~\ref{fig:fbar_cssfr}, we confirm the result of W12 that galaxies with $\Delta C({\rm SF})>0.2$ or $\Delta C({\rm SF})<-0.7$ tend to have higher $f_{\rm bar}$ (fraction of galaxies hosting bars) than the galaxies with intermediate $\Delta C({\rm SF})$. While the prevalence of bars in galaxies with high central SFR has been well accepted \citep{Ellison11, Wang12, CatalanTorrecilla17}, the enhancement of $f_{\rm bar}$ in galaxies with suppressed central SFR has not been studied in detail before, especially in disc-dominated galaxies. The trend can not be explained by the possible dependence of $f_{\rm bar}$ and $\Delta C({\rm SF})$ on $M_*$ or $R_{90}/R_{50}$, for it is steeper than the trend of corresponding control galaxies matched in $M_*$ and $R_{90}/R_{50}$
\footnote{In other words, if the relation between $f_{\rm bar}$ and $\Delta C({\rm SF})$ is caused by both parameters depending on $M_*$ and$\slash$or $R_{90}/R_{50}$, we would observe similar trend of $f_{\rm bar}$ varying as a function of $\Delta C({\rm SF})$ in the parent sample and the control sample. This similarity is not observed, hence there is likely a dependence of $f_{\rm bar}$ on $\Delta C({\rm SF})$ that is independent of  $M_*$ or $R_{90}/R_{50}$. We note that, there might be other unidentified parameters that cause the correlation between $f_{\rm bar}$ and $\Delta C({\rm SF})$. This caveat should be kept in mind and investigated when more abundant observational information becomes available in the future.}. When $\Delta C({\rm SF})<-0.3$, the slopes of linear fits to the relation between $f_{\rm bar}$ and $\Delta C({\rm SF})$ are $-0.21\pm0.07$ and $-0.10\pm0.03$ for the parent sample and the control sample respectively, suggesting the difference of the two trends are significant. 

The results from Fig.~\ref{fig:fbar_cssfr} motivate us to select the following two analysis samples of barred galaxies:
\begin{enumerate}
	\item the {\it LC (Low Central SFR) sample}, with $\Delta C({\rm SF})<-0.7$, in total 49 galaxies.
	\item The {\it HC (High Central SFR) sample}. We select all the galaxies with $\Delta C({\rm SF})>0$ from the barred sample, and match them in the distribution of log $M_*$, $R_{90}/R_{50}$ and redshift z to the LC sample, with bin sizes of 0.15, 0.08, and 0.01 respectively. To ensure the distribution matching to be reliable, we require the K-S (Kolmogorov-Smirnov) test probabilities \citep{Press92} to be above 0.7 (corresponding to a 70\% confidence for similarities) for each of the matching parameters. The resulted HC sample has 48 galaxies. So the HC sample has similar $M_*$ and concentrations (see first row of Fig.~\ref{fig:hist}), but higher (positive, hence enhanced) $\Delta C({\rm SF})$ than the LC sample.
	\item The {\it EHC (Extremely High Central SFR) sample}, with $\Delta C({\rm SF})>0.7$, in total 51 galaxies.
\end{enumerate}

We notice that the LC galaxies on average have higher $M_*$ than the EHC galaxies (panel f of Fig.~\ref{fig:hist}). So in practice, {\bf our analysis will be focused on comparison between the LC and HC samples}, but we also include the EHC sample as a reference for it represents galaxies with the highest $\Delta C({\rm SF})$ in the local universe. 

 We do not exclude the AGN hosting galaxies to avoid the possible bias against AGN hosts. Due to the selection against significant bulges ($R_{90}/R_{50}<2.2$), strong AGN hosts \citep[i.e. Seyferts, identified by the criteria of][]{CidFernandes10} consist only of a small fraction: 8, 2, and 2\% of the LC, HC, and EHC galaxies, respectively. The AGN feedback hence should not play a significant role in affecting the current nuclear SFR. Due to data limitations, the possible influence of a previously active AGN is outside the scope of this paper.

We point out that although the galaxies with the lowest $\Delta C({\rm SF})$ tend to have higher $f_{\rm bar}$ than the galaxies which have intermediate $\Delta C({\rm SF})$, their absolute $f_{\rm bar}$ is low ($\sim$35\%). Also, we could not conclude that bars directly suppress $C({\rm SF})$, i.e., we could not exclude the possibility that an unidentified third parameter causes both the enhancement of $f_{\rm bar}$ and suppression of $C({\rm SF})$ in LC galaxies. This paper hence investigates how bars could be related to the suppression of central SFR in some disc-dominated galaxies while commonly they tend to enhance the central SFR, but does not intend to claim for bars as a major mechanism that quenches star formation in the center of galaxies in general.

\subsection{Other control samples}
\label{sec:control_sample}
There is possibility that the LC sample differs from the HC sample but not from $C({\rm SF})$-normal galaxies. So we also build another control sample (HC2) of 43 galaxies which has $-0.7<\Delta C({\rm SF})<0.7$, and is matched in $M_*$, $R_{90}/R_{50}$, and $z$ to the LC sample like the HC sample. The main purpose of building HC2 is to confirm differences found between LC and HC samples.
We find that the behavior of HC2 are close to that of HC in all trends presented in the following part of this paper. For simplicity, we do not further present or discuss results related to HC2. 

In order to study the radial enhancement$\slash$suppression of SFR in barred galaxies with respect to general galaxies, we build for each of the LC, HC, and EHC samples a control sample that is matched in $M_*$, $\mu_*$, and $g-r$ (referred to as the {\it ($g-r$)-control sample} hereafter). For each galaxy, we select a random galaxy from the parent sample, which differ by no more than 0.1 in $\log M_*$, 0.1 in $\log \Sigma_{*,ct}$, and 0.05 in $g-r$. The K-S test probabilities are all above 0.9 (corresponding to a 90\% confidence for similarities) for each of the matching parameters. We do not specifically select the galaxies without bars, to avoid the danger of being biased against other properties (e.g. low $R_{90}/R_{50}$) favored by the barred galaxies.

\subsection{Cross-matching to other catalogs}
\label{sec:data_othercatalog}
In order to investigate the amount of \hi\ gas in galaxies, we cross-match the LC, HC, and EHC samples with the $\alpha.100$ catalog of the Arecibo Legacy Fast ALFA Survey \citep[ALFALFA,][]{Haynes11}. The nearest ALFALFA counterpart for each galaxy is searched for within a projected distance of 3 arcmin, and a redshift difference of 0.001. 
There are 24, 29, and 24 galaxies covered by the ALFALFA footprint, while 16, 17, and 15 galaxies are detected in the LC, HC, and EHC samples, respectively. The sSFR (SFR$/M_*$) distributions of the ALFALFA detected sub-samples are similar to those of the original samples, with K-S test probabilities of 0.76, 0.96, and 0.99 for the LC, HC, and EHC galaxies, respectively. It suggests that cross-matching with ALFALFA does not result in a significant selection bias toward more star-forming systems.

In order to compare the group environments, we cross-match the three barred samples with the SDSS spectroscopic group catalog of \citet{Lim17}.  \citet{Lim17} identify groups based on a halo-based finder, and use stellar masses as proxy to estimate the dark matter halo mass of each group. There are 33, 35, and 39 of the LC, HC, and EHC galaxies matched to this group catalog, respectively. We take the halo mass ($M_{halo}$) and group richness (number of members in the group, $N_{member}$) for each galaxy from the catalog. 

\subsection{Photometric Radial Profiles}
\label{sec:derive_prof}
We derive the following radial profiles to be analyzed in Sect.~\ref{sec:result_barprof}.
\begin{enumerate}
\item {\bf Ellipticity profiles}. The ellipticity ($e$) of the $r$-band surface brightness isophotes, as a function of the semi-major axis of ellipses fitted to the isophotes. 

\item {\bf Azimuthally averaged profiles}, including the $r$-band surface brightness ($\Sigma_{r,{\rm avg}}$), color ($(g-r)_{\rm avg}$), and stellar mass surface density ($\Sigma_{*,{\rm avg}}$) profiles. The surface brightness is averaged in elliptical rings which have the same axis ratio and position angle as the global values of the galaxy (see Sect.~\ref{sec:parent_sample}). $\Sigma_*$ are derived based on the $r$-band surface brightness, and the $g-r$ dependent $M_*$-to-light ratio \citep{Bell03}. We use the subscript ``ctrl'' to denote measurements for the ($g-r$)-control galaxies. 

\item {\bf Bar profiles:} profiles along the bar, including $(g-r)_{\rm bar}$ and $\Sigma_{*,{\rm bar}}$. The properties are averaged in rectangular grids with width equivalent to the width of the bar ($2\,r_{\rm bar}(1-e_{\rm bar})$) and aligned in the position angle direction of the bar. 

\item {\bf Inter-bar profiles:} profiles perpendicular to the bars, including $(g-r)_{\rm int}$ and $\Sigma_{*,{\rm int}}$. Derived in a similar way as the profiles along the bar, but the grids are aligned in the direction perpendicular to the position angle of the bar. Because $e_{\rm bar}$ have a minimum value of 0.5 in our sample, part of the inter-bar profiles can be contaminated by light from the bar regions when the radius is below 0.5 $r_{\rm bar}$. 
\end{enumerate}

The profiles derived in this paper may suffer from projection effects. The selection of only weakly inclined ($b/a>0.75$) galaxies in the parent sample has mitigated this problem. Considering the potential uncertainties (e.g. thickness of discs and bars, interpolation of data) that might be introduced in deprojection procedures, we only work on these directly derived profiles in this paper. 

\subsection{Azimuthal Fourier Decomposition}
\label{sec:Fourier_decomp}
In order to quantify the amplitudes of non-axisymmetric optical structures, we use Fourier decompositions following the technique outlined in \citet{Yu18}.
The procedure decomposes the $r$-band light along the ellipse at each radius into a series of Fourier components up to an order of six. The ellipticity and position angle of the disk are determined by averaging their profiles in the region where the disc component dominates, and then the image is de-projected to show the face-on disk. The relative amplitude of each Fourier component is calculated as the absolute amplitude over the azimuthally averaged surface density. The relative amplitudes of the $m=2$ component ($A_2$) indicate bar strength within $r_{\rm bar}$ (where the position angle does not significantly change) and two-arm spiral strength beyond $r_{\rm bar}$. 

The cumulative amplitude $S_{A_2,r}$ is calculated as the integral of $A_2$ over the circular area within the radius $r$ of the galaxies. This is also an indicator of bar strength when $r=r_{\rm bar}$, and an indicator of bar strength plus $m=2$ spiral arm strength when $r>r_{\rm bar}$. 

The amplitude $A_{tot}$ is calculated as the square root of the sum of squares of the relative
amplitudes of modes 2, 3, and 4 in the disk-dominated region (beyond $r_{\rm bar}$). $m=3$ and 4 modes are included to account for the multiple-armed structures. $A_{tot}$ hence indicates the non-axisymmetric potential of spiral arms and possibly other non-axisymmetries beyond $r_{\rm bar}$. We use $A_{tot}>0.15$ to indicate a significant existence of spiral arms. 

\subsection{Error bars}
We derive error bars of proportions (fractions) with the formula from \citet{Cameron11}, which estimates the 0.683 (1-$\sigma$) Bayesian binomial confidence intervals from the quantiles of the beta distribution. 

We derive error bars of other statistical quantities (e.g. median values, Kolmogorov-Smirnov (K-S) test probabilities \citealt{Press92}) through bootstrapping. In the bootstrapping procedure, we randomly resample the original sample with replacement, and build 500 new samples each of which has the same size as the original sample. We derive the given statistical quantity $Q$ for each of the 500 new samples, and hence obtain a distribution of $Q$. We calculate the standard deviation of this distribution, and take it as the error estimate of $Q$ of the original sample.

\section{Results}
\label{sec:result}
We compare properties between the LC, HC, and EHC samples in this section.

\begin{figure*} 
\includegraphics[width=15cm]{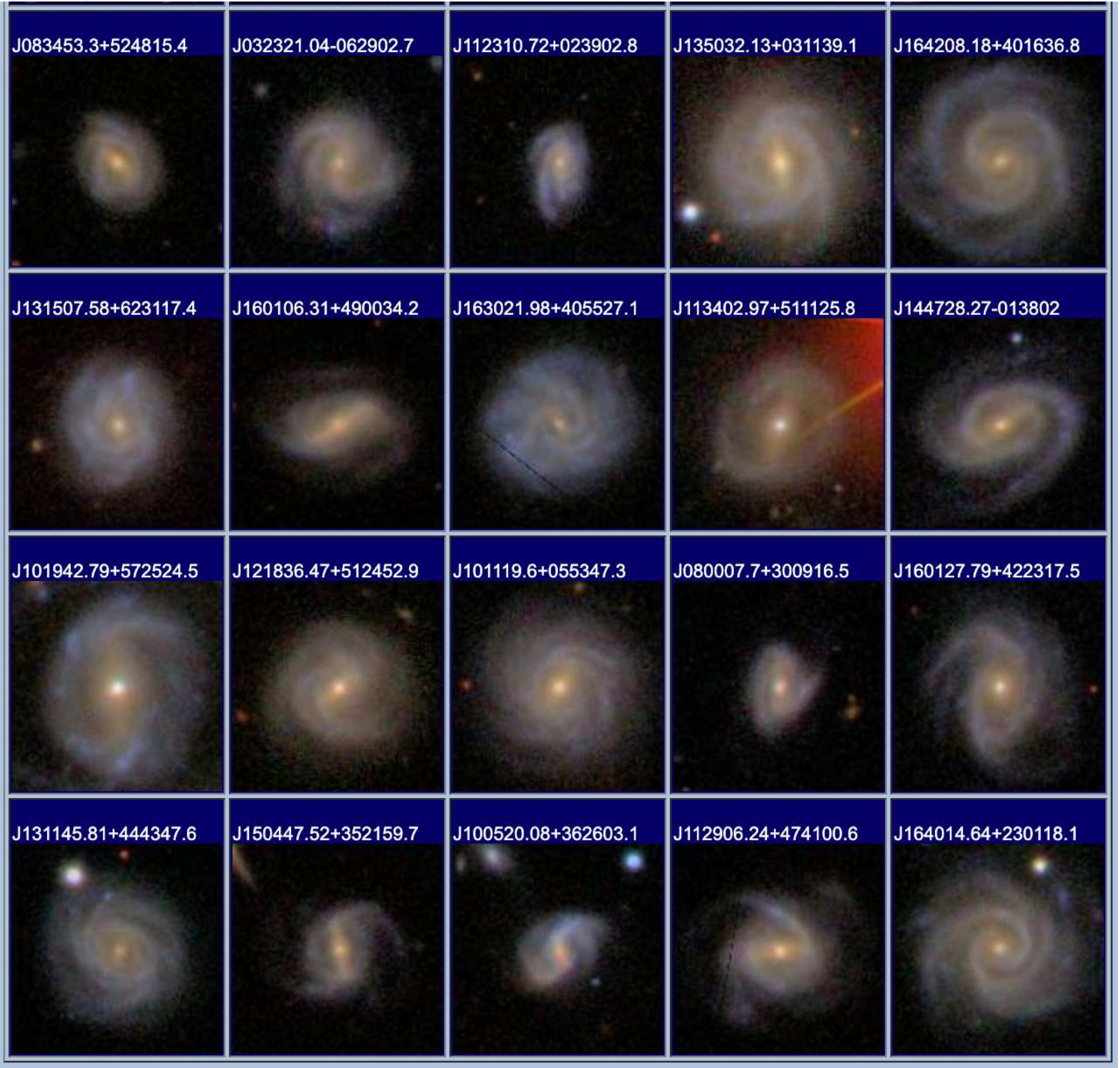}
\caption{An atlas of HC galaxies. The false color images (of 50 arcsec width) are retrieved from the online visual tools of SDSS DR7.}
\label{fig:high_cSF}
\end{figure*}

\begin{figure*} 
\includegraphics[width=15cm]{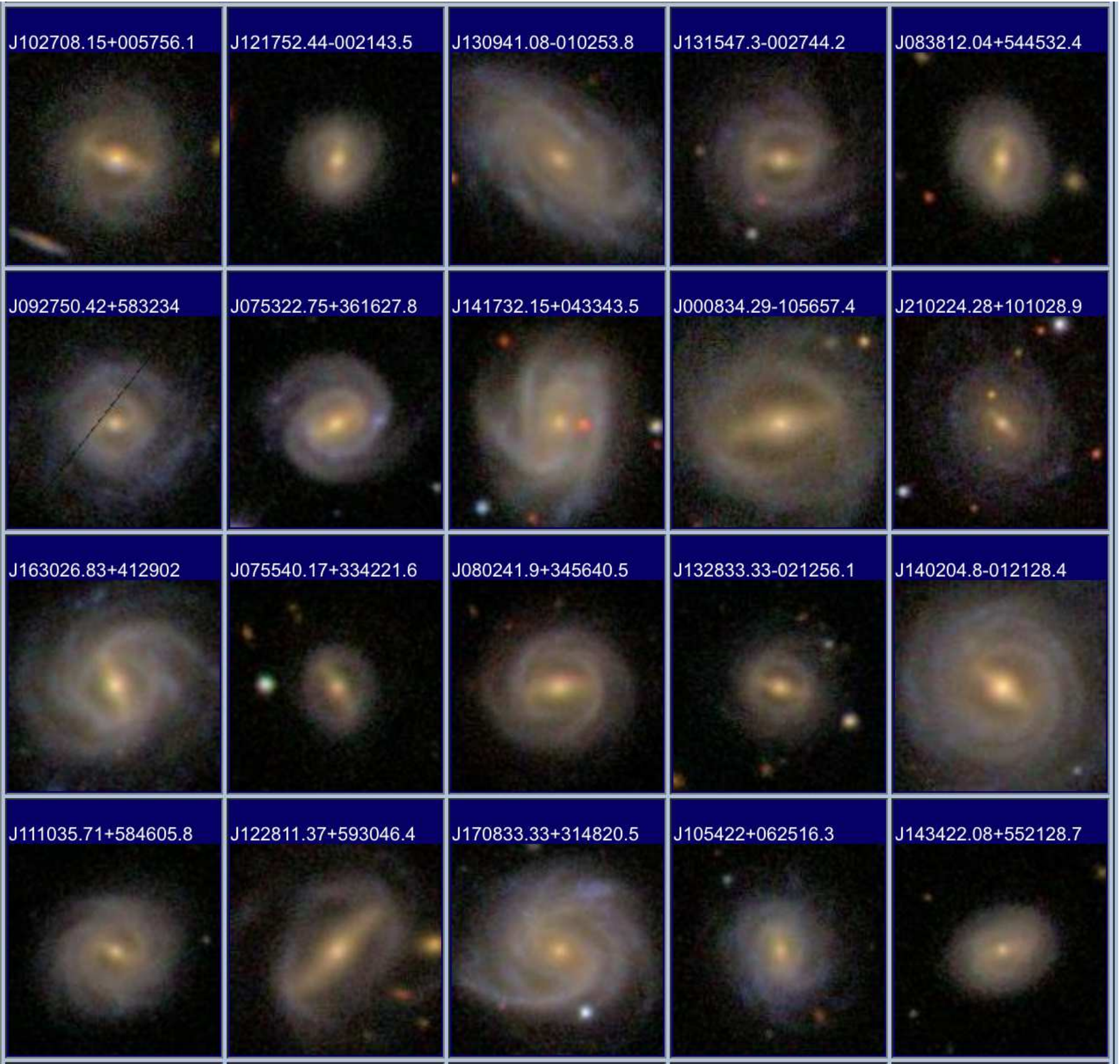}
\caption{An atlas of LC galaxies. The false color images (of 50 arcsec width) are retrieved from the online visual tools of SDSS DR7. }
\label{fig:low_cSF}
\end{figure*}

Fig.~\ref{fig:high_cSF} and \ref{fig:low_cSF} show two atlases\footnote{Obtained from http://cas.sdss.org/dr7/en/tools/chart/list.asp} of examples of galaxies in the HC and LC samples, respectively. Comparing them, we can already see some differences between the morphologies of these two samples. Galaxies in the LC sample look red throughout the galaxy, and very often show closed ring structures around the end of bars. The galaxies in the HC sample often have blue outer discs, and the ends of bars are more likely to connect with spiral arms than with rings. In the following, we will parametrize the properties of these galaxies. 

\subsection{Global properties}
\label{sec:result_global}

We firstly compare the global properties of the whole galaxies. 
\subsubsection{Histograms}

We compare the stellar concentrations, group environments, and global $\hi$-richness of the samples. 

\begin{figure*}
\includegraphics[width=17cm]{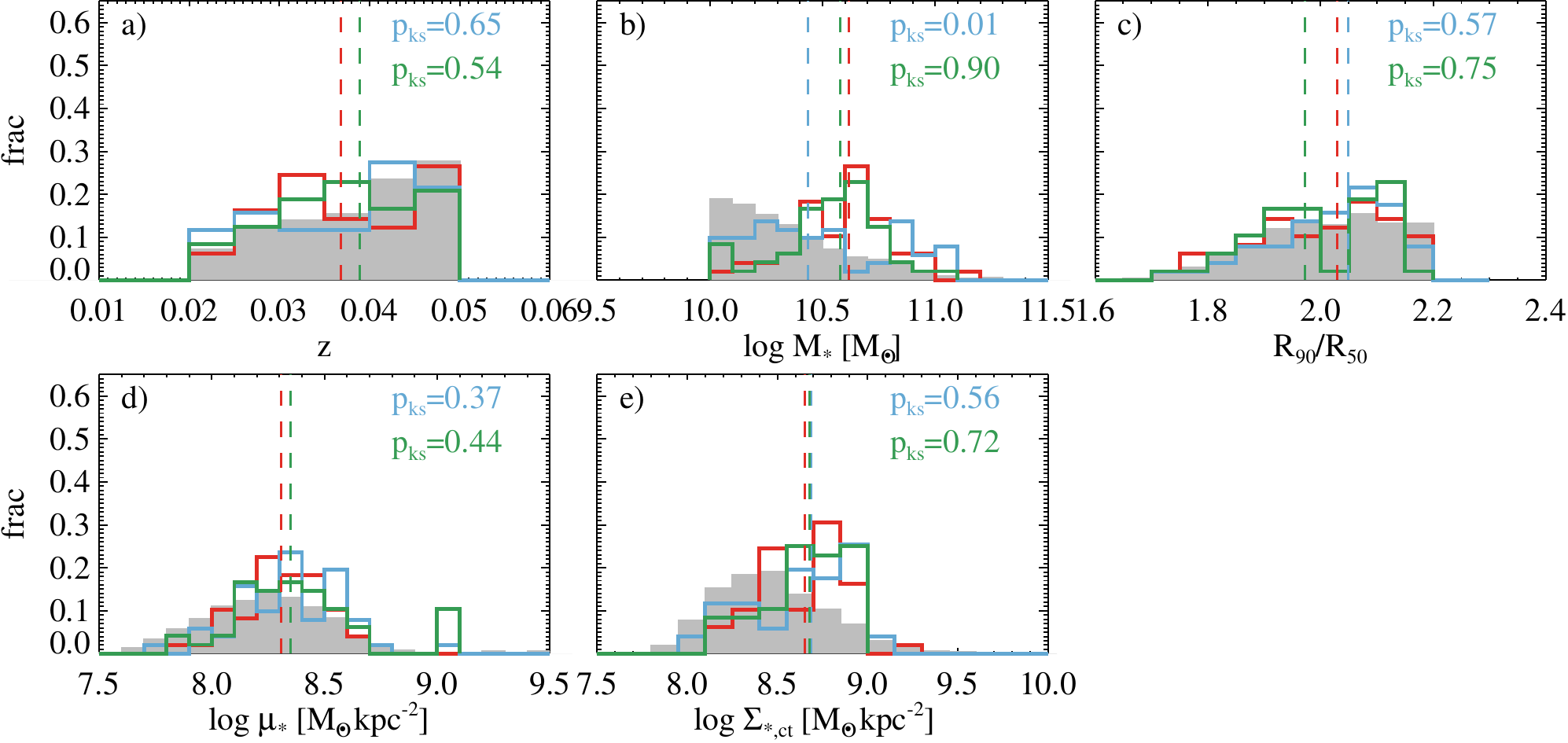}
\includegraphics[width=11.8cm]{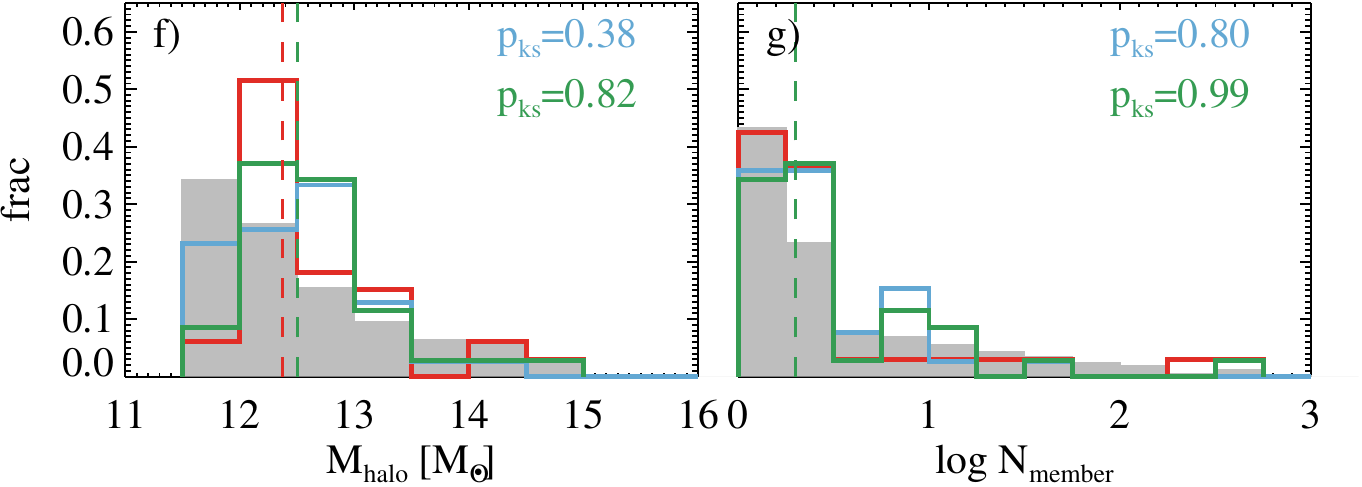}
\hspace{-0.5cm}
\includegraphics[width=5.35cm]{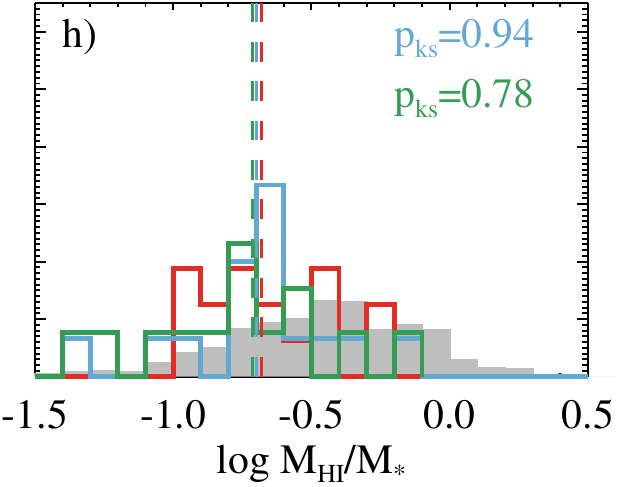}
\vspace{1cm}
\caption{The distribution of global properties of the LC, HC, and EHC galaxies. The first row plots parameters based on which the HC sample is built, including the redshift ($z$), the stellar mass ($M_*$), and optical concentration ($R_{90}/R_{50}$). The second row shows properties related to the inner stellar surface densities, including the effective stellar mass surface densities ($\mu_*$), and the central stellar mass surface densities within a 3-arcsec aperture ($\Sigma_{*,ct}$). Panel f and g in the third row plot properties of the group environment, including the halo mass ($M_{halo}$) and the member richness ($N_{member}$). Panel h plots distributions of the $\hi$ masse fraction ($M_{\mathrm{H\, \textsc{i}}}/M_*$) of the ALFALFA detected galaxies. The grey filled histograms are for the parent sample. The red, green, and blue colors are for the LC, HC, and EHC samples, respectively. The K-S (Kolmogorov-Smirnov) test probabilities \citep[$p_{KS}$,][]{Press92} which indicate the similarity of the distributions between the EHC versus HC samples (blue) and the HC versus LC samples (green) are denoted at the upper-right corner of each panel. The dashed lines show the medians of the distributions.}
\label{fig:hist}
\end{figure*}

SFR surface densities strongly correlate with stellar surface densities in star-forming galaxies \citep{Huang15, Ellison18}.  Central and effective stellar surface densities are also typically used in the literature as a measure of central compactness (bulges), and related to the quenching of central SFR \citep{Woo15}. 
We find that the LC galaxies have similar distributions in the central and effective stellar surface densities as the other two samples (panels c, d, and e of Fig.~\ref{fig:hist}). Hence the LC galaxies are similarly disc-dominated as the HC and EHC galaxies (confirming the selection based on $R_{90}/R_{50}$); the different star-forming status of LC and HC (EHC) galaxies is not likely caused by a difference in the central compactness of stars. 

The group environmental properties, particularly the halo mass is a key parameter determining the gas content and star-forming status of galaxies \citep{Catinella13, Woo15}, for theoretically gas accretion from the circum-galactic medium will be suppressed in in more massive halos \citep{Keres05}. There are 72$_{-9}^{+6}$, 62$_{-9}^{+7}$, and 69$_{-8}^{+6}$\% of the LC, HC, and EHC galaxies identified as central galaxies of groups. The error bars represent the 0.683 (1-$\sigma$) binomial confidence intervals, derived with the procedure of \citet{Cameron11}.
 The LC galaxies have similar distribution of halo mass and group richness (panels f and g of Fig.~\ref{fig:hist}) as the HC and EHC galaxies. Hence the group environments are statistically similar between the three samples. The different star-forming status is not likely caused by the current group environments. 

The star-forming status of galaxies is strongly correlated with the $\hi$-richness \citep{Saintonge16, Saintonge17, Catinella18}. Panel h of Fig.~\ref{fig:hist} shows that the three samples have similar $\hi$ mass fractions. This is consistent with the comparable detection rates of ALFALFA in the three samples ($16/24$, $17/29$, and $15/24$ for the LC, HC, and EHC samples respectively).
Hence it is unlikely that the difference in the central SFR of the three samples is caused by a difference in the mass of $\hi$ reservoirs. 

\subsubsection{Relations}
\label{sec:result_relations}

This section compares the behavior of the samples in a few sequences and scaling relations. 

\begin{figure*}
\includegraphics[width=14cm]{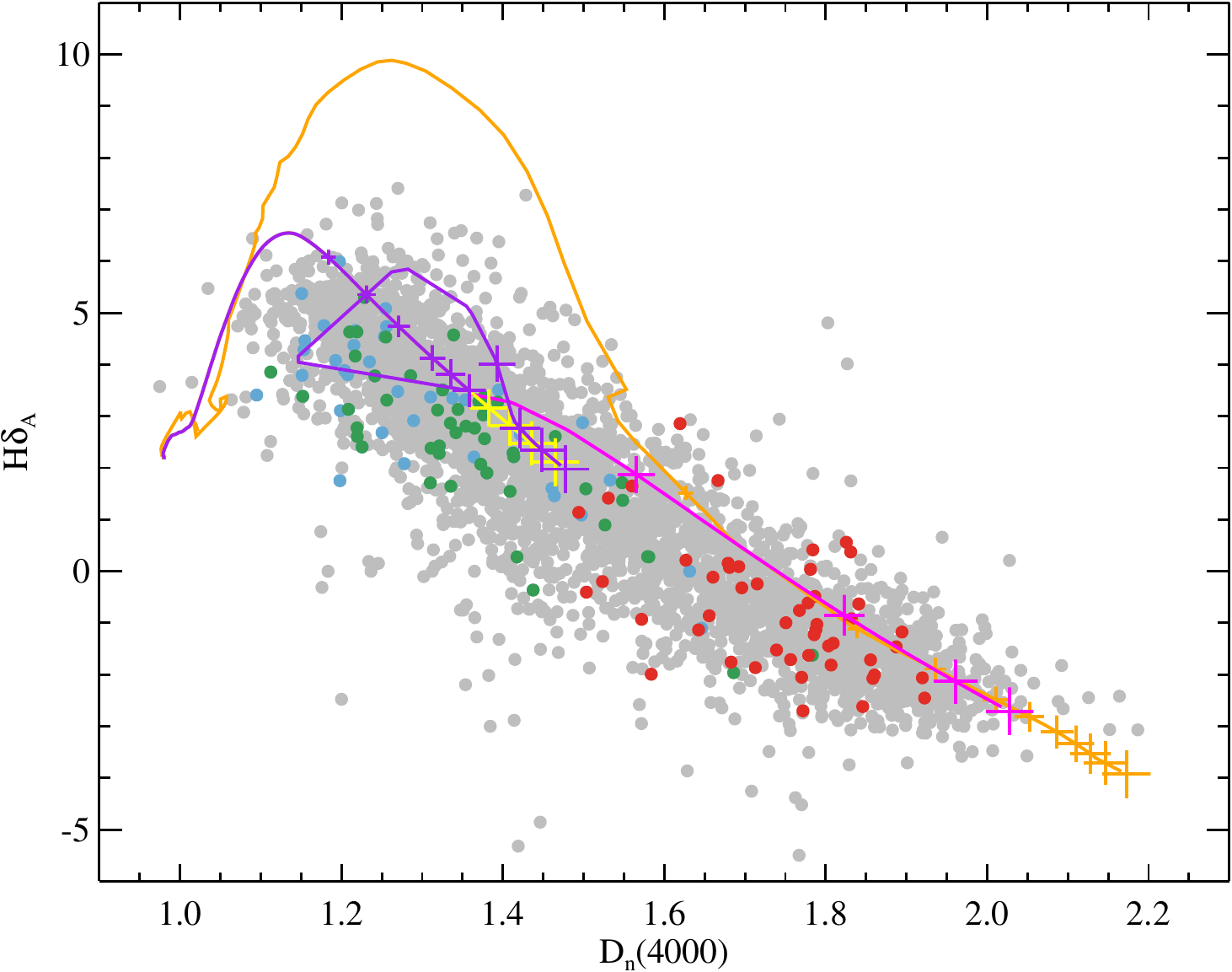}
\caption{ The relation between optical spectra indexes H$\delta_A$ and $D_n$(4000) (see Sec.~\ref{sec:parent_sample}). The grey color is for the parent sample. The curves show the evolution of stellar populations as a function of age with different star formation histories (SFH), obtained with the stellar population synthesis codes of \citet{BC03}, assuming the solar metal abundance. The crosses mark ages of 2, 4, 6, 8, 9, 10, 11, 12, 13, and 14 Gyr on each curve, with larger sizes for older ages. The orange curve is for a single stellar population, and the yellow curve for an exponentially declining SFH, with a characteristic scale-time $\tau$ of 5 Gyr (${\rm SFR}(t)={\rm SFR}(0) \exp{-t/\tau}$). The purple curve is for a single starburst added (at an age of 10 Gyr, lasting for 0.3 Gyr, contributing to 10\% of the total stellar mass formed by 10.3 Gyr) to the SFH of the yellow curve. The magenta curve is for a two-piece wise exponentially declining SFH, with the earlier part (age $<10$ Gyr) the same as the yellow curve, and the later part having a time-scale $\tau$ of 0.5 Gyr (motivated by the observed average property of low-redshift massive galaxies, \citealt{Lian16}).  The yellow, purple, and magenta curves hence overlap when age $<$ 10 Gyr. }
\label{fig:D4000HdA}
\end{figure*}

D$_n$(4000) and H$\delta_A$ are relatively dust-free indicators of the stellar age. 
Star-forming galaxies have lower values of D$_n$(4000) and higher values of H$\delta_A$ than passive galaxies.
Fig.~\ref{fig:D4000HdA} plots the relation between H$\delta_A$ and D$_n$(4000) for the central 3 arcsec of the galaxies. We compare the distribution of galaxies to modeled stellar populations with different SFH. The average position of the HC galaxies in the diagram can be well modeled by adding starbursts to a SFH that is slowly exponentially declining (the purple curve). The single exponentially declining SFH (with or without a starburst) cannot explain the average position of the LC galaxies. Motivated by the average behavior of the low-redshift massive galaxies \citep{Lian16}, if we force the slowly declining exponential SFH to transit at 10 Gyr to a fast declining exponential function (with a time-scale of 0.5 Gyr), then the evolutionary curve reaches the LC galaxies (the magenta curve). Indicated by the magenta curve, the central star formation of the LC galaxies should be quenched more than 1 Gyr ago. The typical stellar age of the LC galaxies should be at least 1 Gyr older than those of the HC (EHC) galaxies if the LC galaxies are evolved from the HC (EHC) galaxies. 

\begin{figure*}
\includegraphics[width=7.8cm ]{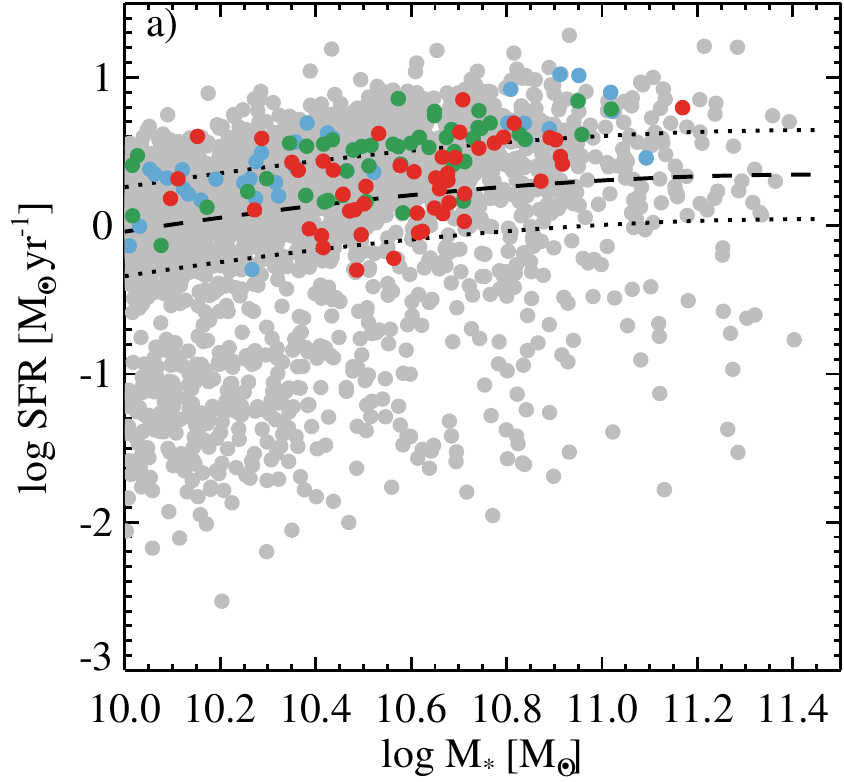}
\includegraphics[width=8.2cm ]{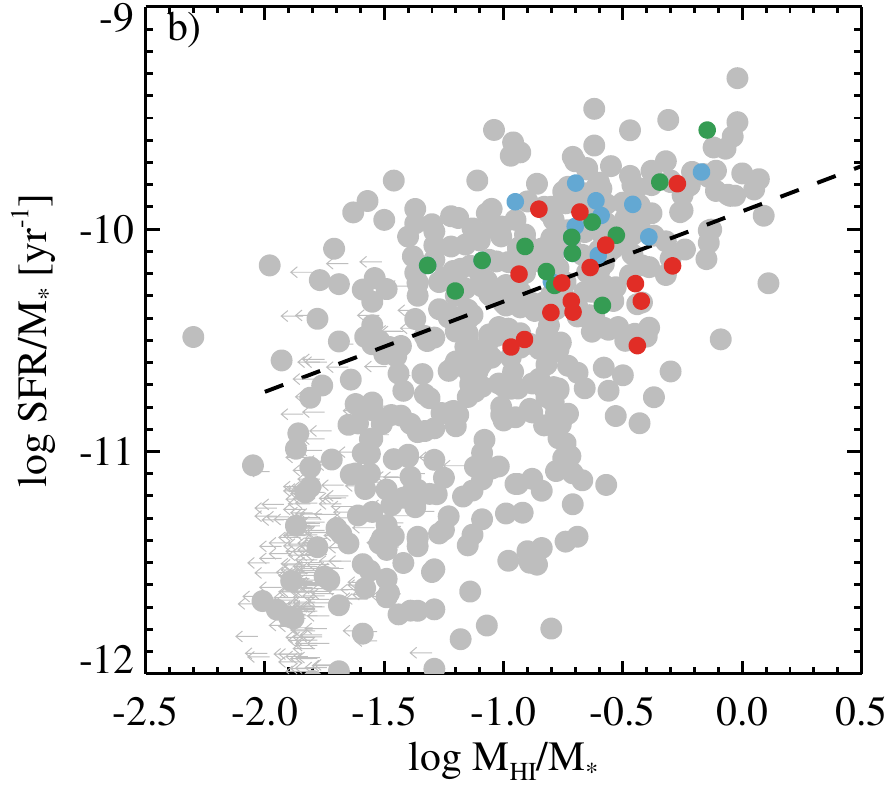}
\includegraphics[width=8cm ]{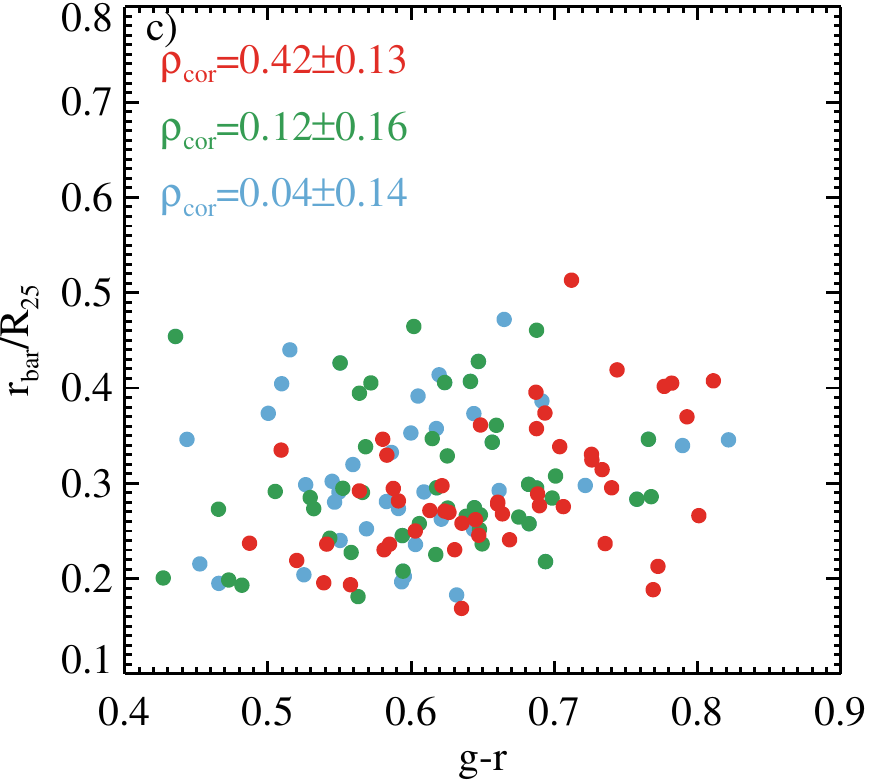}
\caption{Some relations between galactic and bar properties. 
The red, green, and blue colors are for the LC, HC, and EHC samples, respectively. 
{\bf Panel a:} the relation between SFR and $M_*$. The solid and dashed lines mark the mean position and 0.3 dex offset of the SFMS. The grey color is for the parent sample.
{\bf Panel b:} the relation between sSFR and $M_{{\rm HI}}/M_*$. Only the ALFALFA detected LC, HC, and EHC galaxies (see Sec.~\ref{sec:data_othercatalog}) are plotted. The GALEX Arecibo SDSS Survey (GASS, \citealt{Catinella10}) data, which is a $M_*$ and redshift defined $\hi$ sample, is plotted in the grey color, with the detected galaxies shown in dots and the upper limits of the non-detected galaxies shown in arrows. The dashed line shows the robust-fit bilinear relation for the detected galaxies in the GASS sample which have $\log {\rm sSFR}>-11~yr^{-1}$. 
{\bf Panel c:} the relation between the bar length ($r_{\rm bar}$, see Sec.~\ref{sec:bar_sample}) and the global color of the galaxies. The linear Pearson correlation coefficients ($\rho_{cor}$) are denoted, with error bars derived through bootstrapping. }
\label{fig:prop_cor}
\end{figure*}

The parameter space of $M_*$ and SFR is a useful tool in quantifying the global star-forming status of galaxies: the star-forming galaxies show a tight correlation between $SFR$ and $M_*$ (the star-forming main sequence, SFMS) and the quenched galaxies show an extended distribution of low SFR at a fixed $M_*$ \citep{Renzini15}. 
At a fixed $M_*$, galaxies with $SFR$ deviating by less than $\sim$0.3 dex from the median SFMS are considered star-forming, which are possibly fluctuating around the median SFMS in cycles driven by a balance between gas depletion and replenishment \citep{Dekel14, Tacchella16}. We can see from panel a of Fig.~\ref{fig:prop_cor} that the majority of LC galaxies can be classified as star-forming galaxies (i.e. above the curve which is 0.3 dex below the SFMS), though they have on average lower SFR at a fixed $M_*$ than the other two samples. The LC galaxies are not globally SFR-quenched galaxies.

Panel b of Fig.~\ref{fig:prop_cor} plots the relation between sSFR and $M_{\mathrm{H\, \textsc{i}}}/M_*$. On average the LC galaxies have similar $M_{\mathrm{H\, \textsc{i}}}/M_*$ as the HC and EHC samples (with K-S test probabilities of 0.78 and 0.94, respectively). For the same level of $M_{\mathrm{H\, \textsc{i}}}/M_*$, the LC galaxies tend to have lower sSFR than the averaged behavior (the dashed line) of galaxies, while the HC and EHC galaxies are the other way round (with K-S test probabilities of 0.001 and 0.009 respectively, when the distribution of vertical distances from the dashed line in Panel b of Fig.~\ref{fig:prop_cor} is compared to that of the LC sample). It suggests the very low efficiency of LC galaxies compared to other galaxies in converting the $\hi$ gas to SFR.

Theoretical models predict that bars grow in length with increasing age, but gas inflows disturb this process by transferring angular momentum to the bar \citep{Bournaud02, Athanassoula03, Athanassoula13}. We hence expect a correlation between $r_{\rm bar}/R_{25}$ and $g-r$ when there is no significant gas inflow. Panel c of Fig.~\ref{fig:prop_cor} plots the relation between $r_{\rm bar}/R_{25}$ and $g-r$. We find a significant, weak and no correlations in the LC, HC, EHC samples respectively. The higher correlation strength implies the gas inflow rate to be low in the LC sample compared to the HC (EHC) sample.

\subsection{Radial profiles}
\label{sec:result_barprof}
In this section, we compare the distribution of optical light ($M_*$) and color between the different barred samples. 

\subsubsection{Strength of bars and spiral arms}
\label{sec:elprof}
\begin{figure*} 
\includegraphics[width=5.cm]{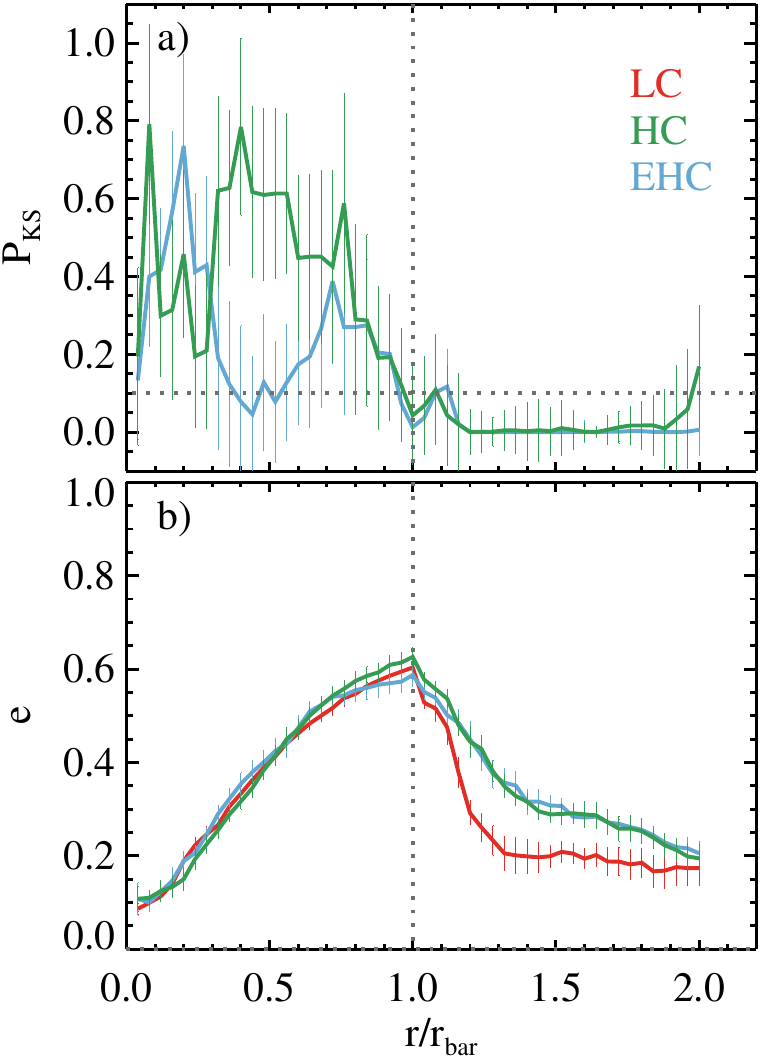}
\hspace{0.5cm}
\includegraphics[width=5.1cm]{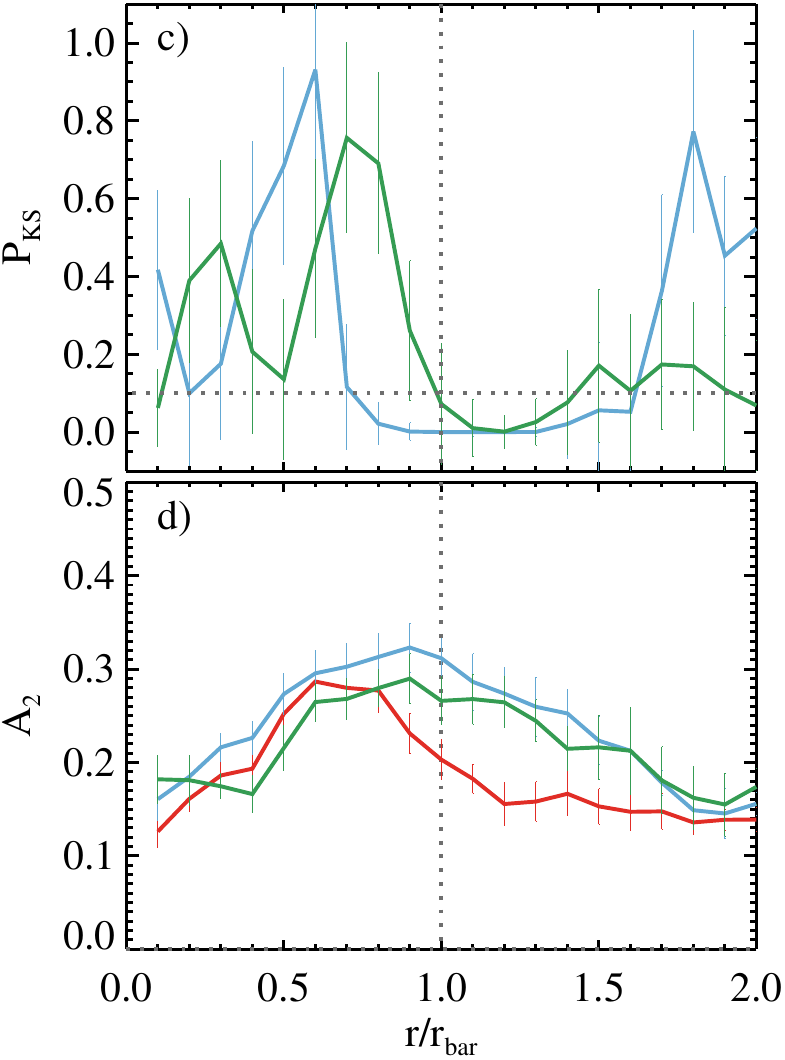}
\hspace{0.5cm}
\includegraphics[width=5.1cm]{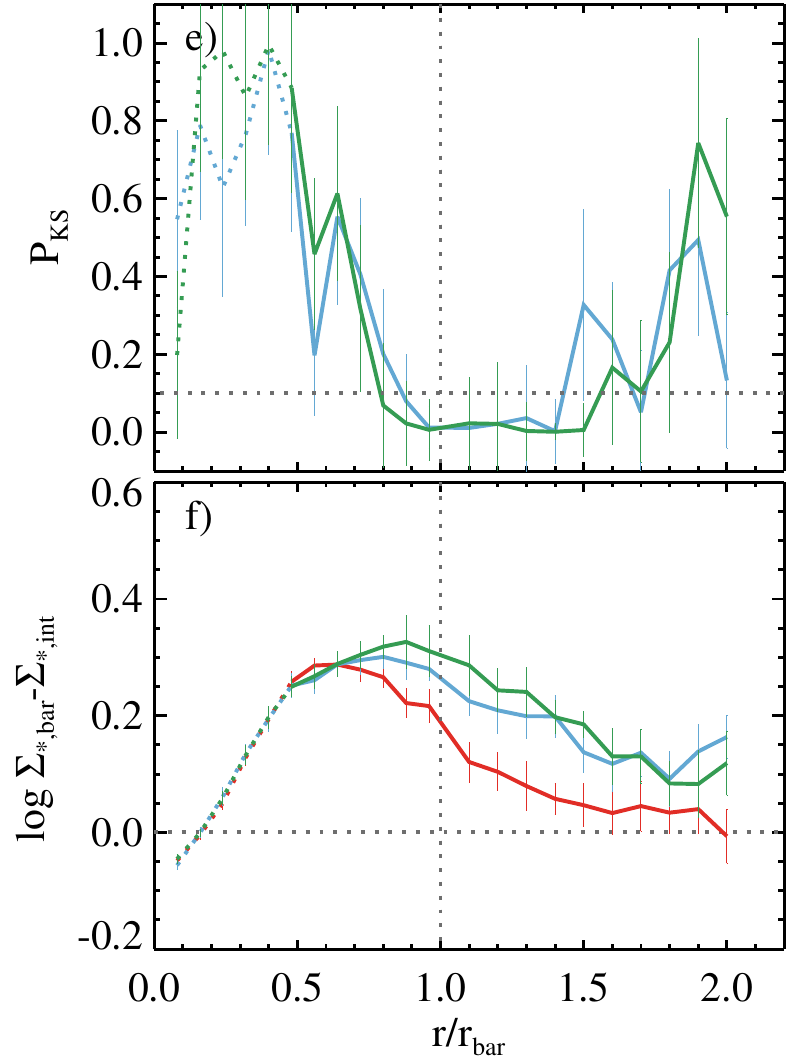}
\caption{ 
Profiles indicating the radial distribution of non-asymmetric structures. The LC, HC, and EHC samples are plotted in red, green and blue, respectively. The left column plots the median profiles of ellipticities ($e$, panel b, see Sec.~\ref{sec:bar_sample} for calculation details) of the three barred samples, and the related K-S test probabilities ($P_{KS}$, panel a) to indicate the similarity of the distribution of the HC$\slash$EHC sample compared to the LC sample at each radius. The middle column plots the median profiles of the relative Fourier amplitudes $A_2$ (panel d, see Sec.~\ref{sec:Fourier_decomp} for calculation details) and related K-S test probabilities (panel c). The right column plots the median profiles of $\Sigma_*$ differences ($\Sigma_{*,{\rm bar}}-\Sigma_{*,{\rm int}}$, panel f, see Sec.~\ref{sec:derive_prof} for calculation details) and related K-S test probabilities (panel e). In panels e and f, the profiles within 0.5 $r_{\rm bar}$ are specifically plotted in colored dotted lines to warn that in this region $\Sigma_{*,{\rm int}}$ can be contaminated by light along the bars. In all panels, error bars are derived through bootstrapping, and the grey vertical lines mark the position of $r_{\rm bar}$. The grey horizontal lines mark 0.1 in panels a, c and e, and 0 in panel f.  }
\label{fig:mst_prof}
\end{figure*}

In Fig.~\ref{fig:mst_prof}, we present radial profiles of three types of parameters that  measure the distribution of the optical light and indicate the strengths of bars and spiral arms. 

In panel b of Fig.~\ref{fig:mst_prof}, we can see that the median $e$ profile of the LC sample looks similar to those of the HC and EHC samples within $r_{\rm bar}$, indicating similar bar strength. The K-S test probability for the comparison between the LC and EHC samples temporarily drops below 0.1 at $\sim0.5\,r_{\rm bar}$, indicating a significant difference in the distribution, despite the similarity of the median values
 of $e$ at that radius. But this difference in $e$ at $\sim0.5\,r_{\rm bar}$ should not be over-interpreted, because the LC and EHC samples have different $M_*$ distributions. 

The median $e$ profile of the LC sample drops more steeply beyond $r_{\rm bar}$ than the other two barred samples. It drops to half the maximum values at $\sim1.2\,r_{\rm bar}$, while the median $e$ profiles of the other two barred samples slowly drops to the same level at $\sim 1.4\,r_{\rm bar}$. The median $e$ profile of the LC sample flattens at $\sim1.3\,r_{\rm bar}$, while those of the other two barred samples continuously drop to the last data points at $2\,r_{\rm bar}$. This is the behaviour one would expect, given the morphology of these galaxies, which was discussed in the beginning of this section. Any strong two-armed spirals in the region beyond the end of the bar will influence the ellipse fit so that the best fit would be far from a circle (see e.g. figure 3 in \citealt{Kalnajs73}). This would be true particularly in the region near the end of the bar, but would also be true considerably further out depending on the strength of the bar. On the contrary, galaxies with an inner ring and no strong spiral beyond it will leave the disc unperturbed except for the regions very near the end of the bar, and have ellipses with much lower ellipticity values. This is indeed what we see in panel b of Fig.~\ref{fig:mst_prof}.

Panel d of Figure~\ref{fig:mst_prof} compares the median radial distribution of  Fourier mode-2 amplitudes ($A_2$) between the samples. LC and HC galaxies have similar $A_2$ along the bars except for the inner-most regions ($r<0.3\,r_{\rm bar}$) and the radius close to $r_{\rm bar}$. Beyond $r_{\rm bar}$, HC galaxies have much higher median $A_2$  than LC galaxies. Hence HC galaxies have similarly strong bars but much stronger spiral arms than LC galaxies.  
EHC galaxies have systematically higher $A_2$ (and $S_{A_2,r_{\rm bar}}$) than the other two samples throughout the radius, however again we refrain from over-interpreting this difference, for EHC galaxies also have systematically lower $M_*$ and only serve as a reference in comparisons between the LC and HC samples.

We show the $\Sigma_*$ contrasts between the bar and inter-bar regions ($\Sigma_{*,{\rm bar}}-\Sigma_{*,{\rm int}}$) in the right column of Fig.~\ref{fig:mst_prof}. From panel f, the median $\Sigma_{*,{\rm bar}}-\Sigma_{*,{\rm int}}$ profiles of the three barred samples are all positive within $r_{\rm bar}$, with a comparable peak value of $\sim$0.3 dex, indicating similar bar strength. But the LC median profile starts to drop from smaller radius and reaches zero (within error bars) at smaller radius than the HC and EHC median profiles. This difference is caused by the fact that the elliptical bar isophotes of the LC galaxies transit into the intrinsically circular isophotes of the outer discs (in many cases related to an inner ring around the bar) much more quickly than those of the HC and EHC galaxies \citep[see][]{Athanassoula02}, and is consistent with the difference found in $e$ and $A_2$ profiles.

\begin{figure*}
\includegraphics[width=16cm]{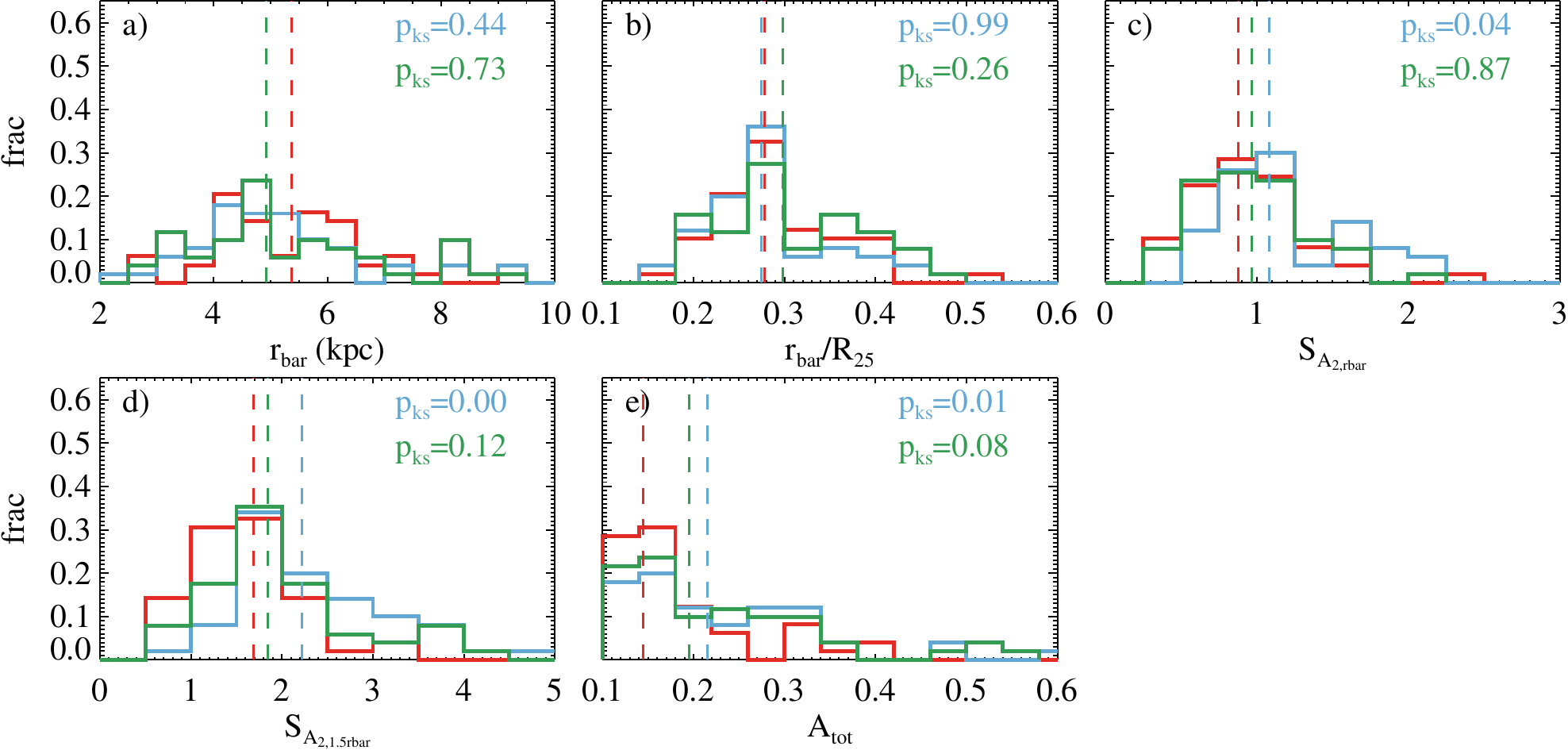}
\caption{ Parameters related to the strength of bars and spiral arms. The parameters include the absolute ($r_{\rm bar}$, panel a) and relative bar lengths ($r_{\rm bar}/R_{25}$, panel b), the radial cumulative relative amplitude of the $m=2$ Fourier decomposition component out to $r_{\rm bar}$ ($S_{A_2,r_{\rm bar}}$, panel c) and $1.5_{\rm bar}$ ($S_{A_2,1.5r_{bar}}$, panel d), and the sum of relative amplitudes of the $m=$2, 3, and 4 components ($A_{tot}$, panel e). Please see Sec.~\ref{sec:bar_sample} and \ref{sec:Fourier_decomp} for details of these parameters. $r_{\rm bar}$, $r_{\rm bar}/R_{25}$ and $S_{A_2,r_{\rm bar}}$ are indicators of the bar strength, and $S_{A_2,1.5r_{\rm bar}}$ and $A_{tot}$ are indicators of the spiral arm strength. The red, green, and blue colors are for the LC, HC, and EHC samples, respectively. The K-S (Kolmogorov-Smirnov) test probabilities \citep[$p_{KS}$,][]{Press92} which indicate the similarity of the distributions between the HC versus LC samples (green) and the EHC versus LC samples (blue) are denoted at the upper-right corner of each panel. The dashed lines show the medians of the distributions. }
\label{fig:hist_bar_spiral}
\end{figure*}

To summarize, the different radial profiles of light$\slash$mass distributions consistently suggest that the LC and HC galaxies have similar bar strength but the HC galaxies have much stronger spiral arms than the LC galaxies. This result is further confirmed in Fig.~\ref{fig:hist_bar_spiral} with a comparison in global measures of the bar and spiral arm strengths. The similar bar lengths (panels a and b) and $S_{A_2,r_{\rm bar}}$ (panel c) confirm the similar bar strengths of LC and HC samples. EHC galaxies show on average higher $S_{A_2,r_{\rm bar}}$ than the other two samples, consistent with the behavior of $A_2$ profiles as discussed above. The on average slightly lower $S_{A_2,1.5r_{\rm bar}}$ (panel d) and significantly lower $A_{tot}$ of the LC galaxies confirm their weaker spiral arms than the HC and EHC galaxies.

\subsubsection{Radial distribution of colors }
\label{sec:result_colorprof}
We use color profiles to indicate the distribution of star forming activities near the bar regions. 

\begin{figure}
\includegraphics[width=8cm]{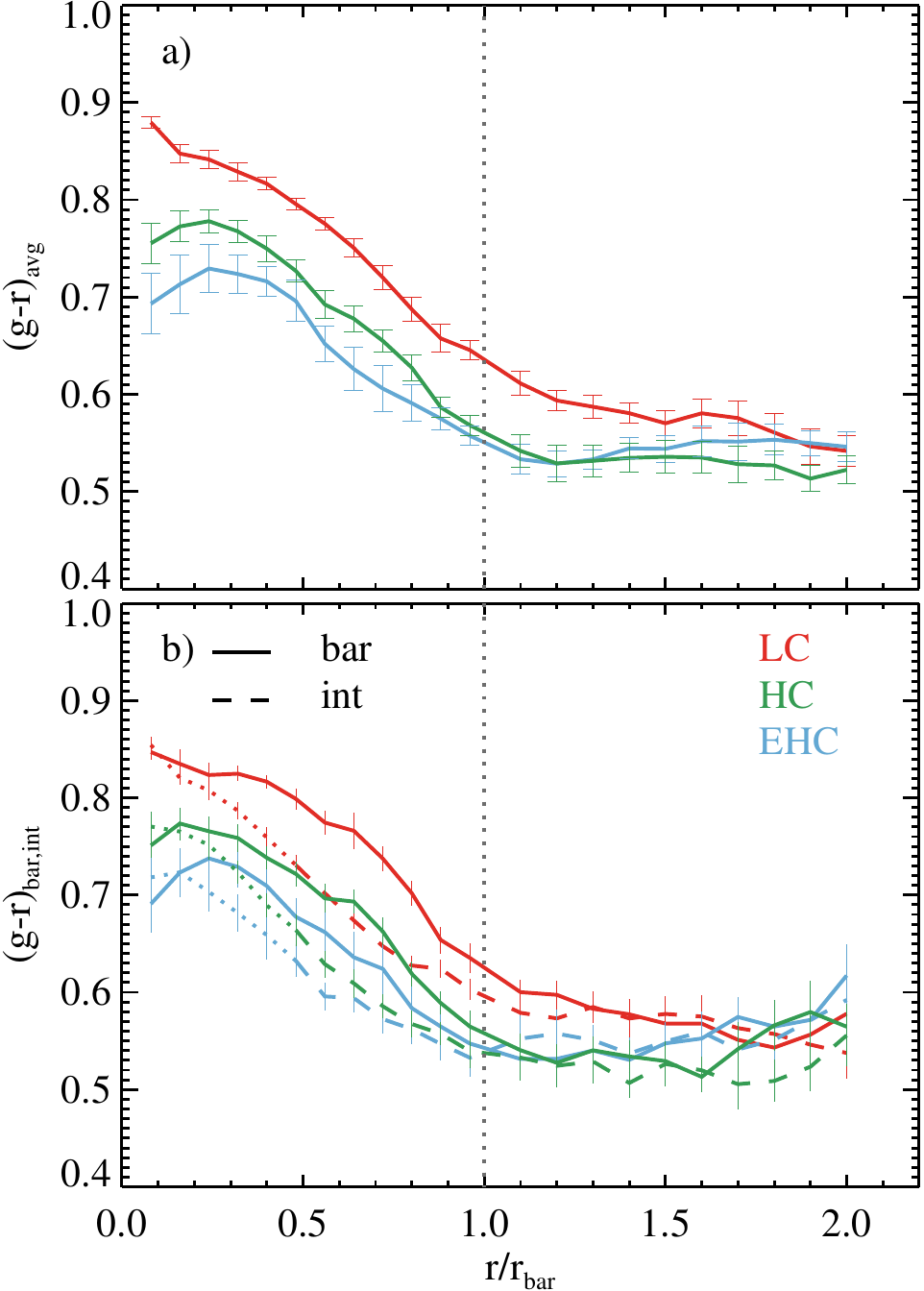}
\caption{  Color profiles for the three barred samples. The LC, HC, and EHC samples are plotted in red, green and blue, respectively.  Panel a plots the median profiles of azimuthally averaged $g-r$. In panel b the median profiles of $g-r$ along the bars (``bar'') are plotted in colored solid lines; median profiles of $g-r$ perpendicular to the bars (``int'') are plotted in colored dashed lines when $r>0.5\,r_{\rm bar}$, and colored dotted lines when $r<0.5\,r_{\rm bar}$.  The $r<0.5\,r_{\rm bar}$ region of the profiles perpendicular to the bars are specifically plotted in colored dotted lines to warn for the possible contamination of light from the bar regions. Please see Sec.~\ref{sec:derive_prof} for details of deriving the color profiles. The error bars of the profiles are calculated through bootstrapping. The vertical grey dotted lines mark the position of $r_{\rm bar}$.   }
\label{fig:gr_prof}
\end{figure}

We show the median profiles of azimuthally averaged $g-r$ (panel a), and $g-r$ along the bars and perpendicular to them (panel b) in Fig.~\ref{fig:gr_prof}. We can see from panels a and b that the median $(g-r)_{\rm avg}$ and $(g-r)_{\rm bar}$ profiles rise toward the center in the LC sample, flattens near the center in the HC sample, and drops near the center in the EHC sample. The difference in color gradients near the center confirms the selection of the three samples based on $\Delta C({\rm SF})$. In panel b, $(g-r)_{\rm int}$ are almost always bluer than $(g-r)_{\rm bar}$ within $r_{\rm bar}$, indicating a common dynamical effect of bars in concentrating or suppressing SFR within the bar radius. 
It also implies that when barred disc-dominated galaxies cease their star formation, it preferentially occurs along the bars (in contrast to simple inside-out or outside-in scenarios).

\begin{figure*}
\includegraphics[width=18cm]{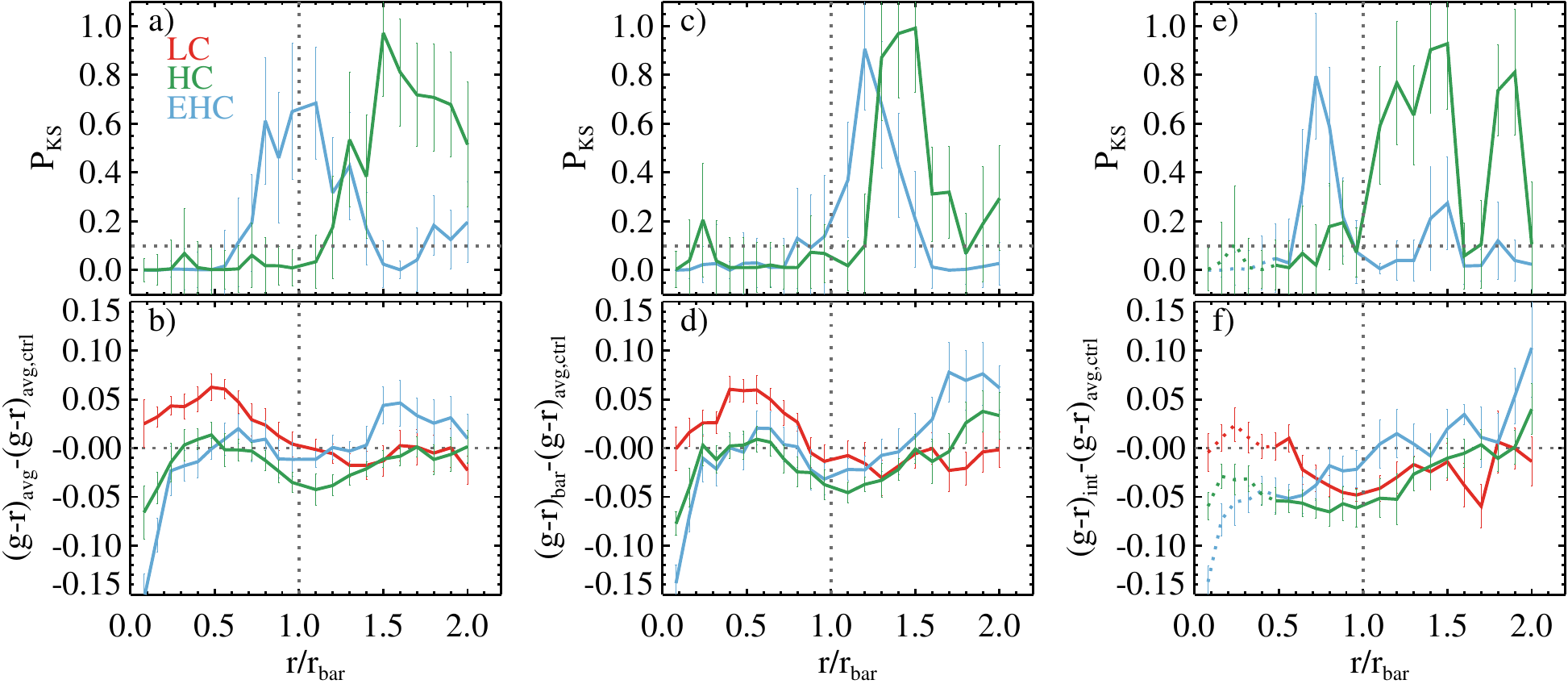}
\caption{ The median profiles of $g-r$ differences from $(g-r)_{\rm avg,ctrl}$ (see Sec.~\ref{sec:derive_prof}) for the three barred samples. The LC, HC, and EHC samples are plotted in red, green and blue, respectively. The left column plots the median profiles of $(g-r)_{\rm avg}-(g-r)_{\rm avg,ctrl}$ (panel b), and related K-S test probabilities ($P_{KS}$) for the HC and EHC samples to be compared to the LC sample in the similarity of distributions at each radius (panel a). The median profiles of $(g-r)_{\rm bar}-(g-r)_{\rm avg,ctrl}$ (panel d) and related K-S test probabilities (panel c) are plotted in the middle column, and the median profiles of $(g-r)_{\rm int}-(g-r)_{\rm avg,ctrl}$ (panel f) and related K-S test probabilities (panel e) are plotted in the right column. The profiles perpendicular to the bars are specifically plotted in dotted lines when $r<0.5\,r_{\rm bar}$, to warn for the possible contamination of light from the bar regions. Error bars calculated through bootstrapping.  }
\label{fig:dgr_ctrl_prof}
\end{figure*}

In Fig.~\ref{fig:dgr_ctrl_prof}, we compare the $(g-r)_{\rm bar}$ profiles of barred galaxies to $(g-r)_{\rm avg,ctrl}$ of the corresponding $(g-r)$-control galaxies (see Sec~\ref{sec:control_sample}). It investigates radially the suppression or boosting of SFR in barred galaxies with respect to the general population of galaxies, i.e. barred versus general (not necessarily barred) galactic internal environments. In panel b, the median $(g-r)_{\rm avg}-(g-r)_{\rm avg,ctrl}$ profiles of the three barred samples differ significantly within $r_{\rm bar}$: the LC profile is positive with a peak value $\sim$0.06 mag within $r_{\rm bar}$, the EHC profile is significantly negative near the galactic center ($<0.2\,r_{\rm bar}$) with a trough value of $\sim-0.15$ mag, and the HC profile is in between of the other two profiles. The outer profiles beyond $r_{\rm bar}$ go the other way from the inner profiles, mostly due to the fact that the barred galaxies and control galaxies are matched in the global $g-r$. 
Similar trends look even clearer along the bars (panel d), but are much weaker perpendicular to the bars (panel f). To summarize, compared to control galaxies with similar global color, the enhancement of SFR in HC and EHC galaxies occur close to the galactic center, while the suppression of SFR in LC galaxies occur within the bar regions.

To summarize, in this sub-section we have shown that the different barred galaxies show similar patterns of SFR distribution around bars; the central SFR is suppressed in LC galaxies not because of peculiar behavior of bar dynamics, but because star formation is suppressed (actually even more seriously) in the whole bar region.

\section{Summary and discussion}
\label{sec:discussion}
\subsection{Similarities and differences between the samples}

We have investigated a sample of massive, disc-dominated ($R_{90}/R_{50}<2.2$), barred galaxies which have suppressed central SFR (LC), by comparing them to barred galaxies which have significantly enhanced central SFR (HC and EHC). 

 We found the following common features for these three barred samples, which implies that the differences between their (central) SFR in galaxies cannot be due to them.

\begin{enumerate}[label=(\roman*)]
\item {\bf Stellar central compactness}. The galaxies in these three sub-samples have similar distributions of stellar mass central surface densities and optical concentrations (panels c, d, and e of Fig.~\ref{fig:hist}).

\item {\bf Group environment}. They have similar distributions of group masses and group richness (panels f and g of Fig.~\ref{fig:hist}).

\item {\bf HI gas abundance}. They have comparable ALFAFLA detection rate, and similar $M_{\mathrm{H\, \textsc{i}}}/M_*$ (panel h of Fig.~\ref{fig:hist} and panel c of Fig.~\ref{fig:prop_cor}).

\item {\bf Bar strength}. The HC and LC galaxies have similar bar strengths measured radially (Fig.~\ref{fig:mst_prof}), and globally (panels a, b,  and c of Fig.~\ref{fig:hist_bar_spiral}).

\item {\bf Distribution of SFR close to bars}. The regions along the bar are redder than the inter-bar regions (Fig.~\ref{fig:gr_prof}), and the central star formation is more enhanced (less quenched) with respect to control galaxies (indicated by $(g-r)-(g-r)_{ctrl}$) than the star formation in main bar regions (Fig.~\ref{fig:dgr_ctrl_prof}). 
\end{enumerate} 

There are three major differences found between the LC and HC (EHC) samples, which are keys to understanding the suppressed central SFR in the LC galaxies.
\begin{enumerate}[label=(\roman*)]
\item {\bf Bar length related to color}. LC galaxies have stronger correlation between relative bar sizes and global colors than the other samples (panel d of Fig.~\ref{fig:prop_cor}).

\item {\bf $g-r$ color out to 2 $r_{\rm bar}$}. The LC galaxies on average have redder disks than the HC (EHC) galaxies through out the radius range (Fig.~\ref{fig:gr_prof}). The difference gets smaller when $r>1.5\,r_{\rm bar}$.

\item  {\bf Spiral arm strength}. The LC galaxies have much weaker spiral structures beyond the bar radius (Fig.~\ref{fig:high_cSF}, \ref{fig:low_cSF}, and \ref{fig:mst_prof}, and panels d and e of Fig.~\ref{fig:hist_bar_spiral}). 
\end{enumerate}

Before we discuss possible scenarios, we note that the differences observed between LC and HC (EHC) galaxies can easily be confused with the differences known between early-type and late-type galaxies. Early-type galaxies tend to be redder, to show weaker spiral arms, and to have rings more frequently (\citealt{Buta15}, and references therein)  than late-type galaxies. Early-type galaxies, however, are also expected to have more prominent classical bulges (higher central surface densities and higher central concentrations in the stellar distributions) and lower neutral gas content, while both LC and HC (EHC) galaxies are similarly (and highly) disc-dominated and $\hi$-rich. 
We hence cannot just attribute the difference between LC and HC (EHC) galaxies to a simple difference between different Hubble types. 

\subsection{Indications from the similarities}
There are a few scenarios commonly related to star formation cessation but we can consider them as unlikely because of the similarities of the LC and HC samples.  

\subsubsection{ Star formation cessation due to stellar compactness and environments}
Observations and theories suggest that compact centers and massive halos are two key conditions for galaxies to cease their central star formation \citep{Woo15, Dekel14}.  The combination of these two factors is often discussed under the context of the ``compaction model'', where star-forming galaxies evolve around the star forming main sequence, regulated by a balance between depletion and replenishment of the neutral gas \citep{Tacchella16}. The compact centers could be related to feedback from starbursts or the black hole, which quickly deplete or remove the central gas, while the massive, hot halo prevents accretion of the neutral gas \citep{Zolotov15, Woo15}.
The similarities in related properties suggest that it is unlikely that these two facts are responsible for the SFR-suppressed center of LC galaxies. 

\subsubsection{ Star formation cessation due to global HI abundance }

Although molecular gas is more directly material or tracer for star formation than $\hi$ gas, its mass is typically a relatively small fraction ($\sim20\%$) of the neutral gas, and its depletion time is $\sim$1 Gyr for galaxies close to the star-forming main sequence \citep{Saintonge11, Saintonge17, Catinella18}. Hence the $\hi$ gas is considered as a necessary reservoir to sustain the star-forming status in galaxies. It has been confirmed by previous studies that SFR is strongly correlated with $\hi$ mass fraction at a given stellar mass \citep{Saintonge16, Saintonge17}.
The similarity in $\hi$ mass fractions between the LC and HC (EHC) samples suggests that the global $\hi$-richness is unlikely to be the cause of suppressed central SFR in the LC galaxies. It is more likely that the large $\hi$ reservoir of the LC galaxies for some reason could not efficiently flow in to fuel the star formation within the (relatively) inner disks.

\subsubsection{Star formation cessation due to local star forming efficiency of HI gas}

A very low star forming efficiency may cause a similarly low SFR, even when plenty of HI gas
is present. Star formation models predict that the localized star forming efficiency strongly depends on the stellar mass surface density and metallicity \citep{Ostriker10, Krumholz09, Krumholz13}. The LC, HC, and EHC samples have similar $\mu_*$ and $\Sigma_{*,ct}$, and the localized metallicities are strongly correlated with the stellar mass surface densities \citep{Carton15}. Hence on average, these three samples are unlikely to have different central star forming efficiencies due to different metallicities or stellar mass surface densities. 

 \citet{Krumholz15} predicted that strong turbulence caused by gas inflows could temporarily reduce the central star forming efficiency in barred galaxies until the gas cumulates in the center to very high densities; this cycle runs on a typical timescale of a few tens Myrs. However, the centers of the LC galaxies are on average older than the HC and EHC galaxies by $\ga$1 Gyr, and the short-period fluctuations of SFR predicted by \citet{Krumholz15} is likely to be averaged out on such long time-scales. Hence the scenario of \citet{Krumholz15} is unlikely to explain the formation of the LC galaxies.

\subsubsection{ Star formation cessation due to bar strength}
\label{sec:discuss_bar}

Both observations \citep[e.g.][]{Sakamoto99, Sheth05, Wang12}
and simulations (A92b; \citealp{Piner95, Athanassoula13, Sormani15}, etc.) have shown  that gas in the region 
between CR and ILR moves inwards due to the bar torques.
The same is true for stars, albeit to a much lesser extent \citep[e.g.][]{Athanassoula02, Valenzuela03}. 
If the region within ILR (i.e. the central region) has non-axisymmetric features -- like inner bars, ovals, or  
spirals -- material that has been pushed inwards to the ILR can be pushed further inwards, yet closer
to the centre \citep[][etc.]{Shlosman89, Peeples06, Schinnerer06, Meier08, Cole14}. In the region outside CR the direction of the radial motion has the opposite sign, i.e. is outwards \citep{Kalnajs78} and gas does not cross CR. 

Thus, the gas initially within CR is accumulated in the central
region, while the region between ILR and  CR is steadily depleted of its gas,
except for two narrow stripes along the leading sides of the bar, which
are the shock loci (A92b). Thus, there should be considerable star formation in the central region, but practically none further out in the main bar region, because the shear in the
shock loci prevents star formation even in the narrow, high density regions there,
provided the bar is sufficiently strong \citep[A92b;][]{Sorai12, Meidt13,George19}. 
A92b (see figures 6 and 7, and section 4.2 there) showed that stronger bars push more material
inwards and also create more extended and
emptier regions between ILR and CR and more gas concentration in the
centre \citep[for the latter see][]{Athanassoula94, Athanassoula13}. It would thus be natural to associate stronger bars with more centrally concentrated SF.  

The LC and HC (EHC) samples show similar SFR distribution near the bar regions (i.e. in the comparisons of bar versus inter-bar regions, and main bar versus central regions), consistent with the pattern produced by gas inflows along strong bars, as described above. They also have similar bar strength. Hence,  the suppressed central SFR in the LC galaxies is unlikely to be caused by bars of different strength inducing different gas inflows.

\subsection{Indications from the differences: why is star formation centrally concentrated in some barred galaxies and not in others?}
The different properties between the samples are possibly related to the processes that cause the different central SFR.
The strong correlation of relative bar lengths and global color in the LC sample is consistent with the theoretical prediction
that the bar is longer in cases with no gas, or little gas, than in gas-rich cases (\citealt{Athanassoula13}, particularly figures 4 and 5 there, and \citealt{Athanassoula14}), and consistent with our speculation above, based on the $\hi$ abundance, that gas inflows should be weak in the LC galaxies. 
Below we discuss a possible cause for LC galaxies to have weak gas inflows.

\subsubsection{A possible scenario related to spiral arms}
The most prominent difference between the samples is in
the strength of the spiral arms. The dynamic relation between
bars and spiral arms can be complex and one can distinguish two
possibilities. Some theories, such as simple bar driven spirals \citep[e.g][]{Athanassoula78}, or the standard manifold theories \citep[][etc.]{RomeroGomez06, RomeroGomez07,
 Voglis06, Patsis06, Athanassoula10, Athanassoula12,
Efthymiopoulos19} assume that the pattern speed
($\Omega_p$) of the spirals is equal to that of the bar and thus that
the spirals are outside their corotation
radius (CR). In such cases, the radial motion of the gas
will be outwards \citep{Kalnajs78}, so there will be no enhancement
of the central SFR due to spiral arms.

This assumption, however, need not be true for all barred galaxies (see
e.g. discussion in sect. 6.4 of \citealt{Athanassoula10}), and in
some cases the bar and spiral may have different pattern speeds and yet
exchange energy and angular momentum between them, provided certain
conditions are met and two resonances, one from
each component, overlap sufficiently \citep{Tagger87, Sygnet88}. In such cases, the spiral continually breaks from and
reconnects to the bar, providing a realistic morphology \citep{Sellwood88}. A detailed study by \citet{Masset97} argued
that such a non-linear coupling between the ILR of the spiral and the
CR of the bar can be quite efficient, and even more relevant than the
swing mechanism in accounting for the dynamics of the galaxy beyond the
corotation of the bar. Thus the spiral is within its own CR and the
gas in that region can be pushed inwards and therefore produce a
larger gas concentration than what would have been achieved by the bar
on its own. Thus, the much
stronger spiral arms in the HC (EHC) galaxies would help explain
the difference between the central SFR of LC and HC/EHC galaxies.

The second main difference, namely the color, is naturally linked to the
spiral strength. Indeed for a given forcing, i.e. in this case a given
bar strength, the resulting spiral would be stronger if the stellar
population constituting the outer region was kinematically colder (see e.g. \citealt{Athanassoula84}, for a review), i.e. younger and therefore bluer.

\section{Conclusion and future perspective}
\label{sec:conclusion}
In this paper we compared two types of barred galaxies, those with centrally suppressed SFR (LC) and those with centrally enhanced SFR (HC$\slash$EHC). By making averages over each group, we compared various global quantities related to their mass, density, gas fraction, SF, colors and bar properties, as well as various radial profiles. We found which features are common between the two groups and which are clearly different. In the latter let us mention the star forming status in the center, inner disks, and the spiral arm strength, while the former includes group environment, stellar density radial distributions, global HI richness etc. We searched for clues under the context of galaxy dynamic theories and simulations. After eliminating a number of alternatives, we proposed one possible scenarios to produce the major differences of LC and HC (EHC) galaxies, while preserving most of the similarities. This scenario relies on the fact that only gas within the CR can be driven inward by non-axisymmetric components, such as bars or spirals. The spirals have lower pattern speeds than the bars, while being coupled to them via resonances. In such cases, both the spirals and the bar could be driving gas inwards, thus increasing the amount of gas driven to the center of HC galaxies. 

Our paper contributes considerable new input on the problem at hand, but is  far from having solved it. We briefly summarize here some possible improvements.
 
This paper relies on analysis of radial profiles, but does not include a decomposition of the light into different components. The selection of low-$R_{90}/R_{50}$ galaxies may have mitigated the contamination from a central bulge, however, the light from the disc may hide structures in the bars, and disc breaks may also affect our analysis. Analysis based on individual components obtained from decompositions, such as achieved by \citet{Kim15, Kim16}, \citet{Salo15}, and \citet{Gao17} will be helpful and important to confirm the results in this paper. The decompositions will also enable  analysis for galaxies with higher $R_{90}/R_{50}$. 

Some of our results need confirmation from other types of data in the future. Optical or near infrared spectroscopy either based on long-slit or IFU (integral field unit, MaNGA, \citealt{Bundy15}, for example) equipments will enable us to obtain mean velocities as well as velocity dispersions of the stellar component, which, in turn, may allow us to get further information (e.g. pattern speed, corotation, kinematical temperature) on the dynamical evolution of our bars. The new radio instruments (e.g. MeerKAT and ALMA) will be useful not only to directly map the distribution of cold gas, but also to check whether the radial variation of SFE is the same among the different types of galaxies. Kinematical analysis of the cold and ionized gas may directly reveal the inflow and outflow of gas.  

Direct comparisons of observations to simulations like those in \citet{Athanassoula13, Athanassoula16} will help us gain more insight into the physics that produces the observables. The most important improvement, however, would be to introduce results not from a single snapshot but from a number of them, all with appropriate properties. Then averages could be taken over all corresponding simulated galaxies, as we do here with the observed galaxies in our sample. We also need to study in depth the SFR in each of these, in order to make more specific comparisons. These will be subjects of future research.

\acknowledgments
We gratefully thank M. Krumholz, K. Wang, S. Ellison for useful discussions.

E. Athanassoula thanks the CNES for
financial support. This work was granted access to the HPC
resources of CINES under the allocation 2019-A0060407665
and 2018-A0040407665 attributed by GENCI (Grand Equipement
National de Calcul Intensif). This work was granted access to the HPC resources of Aix-Marseille Université financed by the project Equip@Meso (ANR-10-EQPX-29-01) of the program Investissements d'Avenir supervised by the Agence Nationale de la Recherche.

LS thanks the support by the China Postdoctoral Science Foundation (2018M641067).

GALEX (Galaxy Evolution Explorer) is a NASA Small Explorer,
launched in April 2003, developed in cooperation with the
Centre National d'Etudes Spatiales of France and the Korean Ministry
of Science and Technology.

We thank the many members of the ALFALFA team who
have contributed to the acquisition and processing of the ALFALFA
dataset over the last many years. RG and MPH are supported by NSF
grant AST-0607007 and by a grant from the Brinson Foundation.

Funding for the SDSS and SDSS-II has been provided by
the Alfred P. Sloan Foundation, the Participating Institutions, the
National Science Foundation, the U.S. Department of Energy,
the National Aeronautics and Space Administration, the Japanese
Monbukagakusho, the Max Planck Society, and the Higher Education
Funding Council for England. The SDSS Web Site is
http://www.sdss.org/.

\bibliographystyle{apj}
\bibliography{cSFRbar}

\appendix
\section{Parameters and abbreviations for terms used in the main part of the paper}
In table~\ref{tab:para}, we list the abbreviations for terms and parameters used in this paper. The terms and parameters of each category are listed following the alphabet order.
 We list for each abbreviation or parameter in the table a short description, and reference to the section in the main part of the paper where the abbreviation or parameter was defined. 

\begin{table}
\centering
\caption{ Parameters and abbreviations for terms  }
{
\begin{tabular}{c l c }
Definition  &   description  &  section for definition \\

\hline
{\bf Datasets:} & & \\
ALFALFA & Arecibo Legacy Fast ALFA Survey &\ref{sec:parent_sample}\\ 
GALEX & Galaxy Evolution Explorer &\ref{sec:parent_sample}\\ 
GASS & The GALEX Arecibo SDSS Survey &   \ref{sec:result_relations} (Fig.~\ref{fig:prop_cor})\\ 
MPA$\slash$JHU & An advanced parameter catalog for SDSS & \ref{sec:parent_sample}\\ 
SDSS &  Sloan Digital Sky Survey &\ref{sec:parent_sample}\\ 

\hline
{\bf Galactic abbreviations:} & & \\
CR  & corotation  &  \ref{sec:introduction} \\
ILR & Inner Lindblad Resonance &   \ref{sec:introduction}  \\
SED & spectral energy distribution & \ref{sec:parent_sample}\\ 
SFH & star formation history  & \ref{sec:parent_sample}\\ 
SFMS & star forming main sequence & \ref{sec:result_relations}\\

\hline
{\bf Galactic properties:} & &  \\
$A_2$ & a measure of bar strength and arm strength & \ref{sec:Fourier_decomp} \\
$A_{tot}$ & a measure of spiral arm strength  & \ref{sec:Fourier_decomp} \\
$C({\rm SF})$  & central concentration of specific SFR & \ref{sec:parent_sample}\\ 
$\Delta C({\rm SF})$  & enhancement of C(SF) with respect to control galaxies & \ref{sec:parent_sample}\\ 
$D_n(4000)$ &  4000-$\angstrom$ break  & \ref{sec:parent_sample}\\ 
$e$ & ellipticity   & \ref{sec:bar_sample} \\ 
$e_{\rm bar}$ & maximum ellipticity of the bar & \ref{sec:bar_sample} \\ 
$f_{\rm bar}$ & fraction of galaxies hosting strong bars & \ref{sec:subsample}\\ 
$g-r$  & optical color   & \ref{sec:parent_sample}\\ 
$(g-r)_{\rm avg}$ &  azimuthally averaged $g-r$ profile  & \ref{sec:derive_prof} \\
$(g-r)_{\rm bar}\backslash\Sigma_{*,{\rm bar}}$ & radial distribution of $g-r\backslash$stellar surface density perpendicular along the bar  & \ref{sec:derive_prof} \\
$(g-r)_{\rm int}\backslash\Sigma_{*,{\rm int}}$ & radial distribution of $g-r\backslash$stellar surface density perpendicular to the bar & \ref{sec:derive_prof} \\
$H\delta_A$ &   Balmer absorption index & \ref{sec:parent_sample}\\ 
$M_*$ & stellar mass & \ref{sec:parent_sample}\\
$M_{halo}$ & halo mass & \ref{sec:data_othercatalog} \\
$M_{\mathrm{H\, \textsc{i}}}$ & $\hi$ mass & \ref{sec:data_othercatalog} \\
$\mu_*$ & effective stellar mass surface densities & \ref{sec:parent_sample}\\ 
$N_{member}$ & number of galaxies in the group & \ref{sec:data_othercatalog} \\
$R_{90}/R_{50}$ & central concentration of light   & \ref{sec:parent_sample}\\ 
$R_{25}$ & optical disk size   & \ref{sec:parent_sample}\\ 
$r_{\rm bar}$ & bar length & \ref{sec:bar_sample}\\
$S_{A_2,(1.5)r_{\rm bar}}$ & a measure of bar (and spiral arm) strength & \ref{sec:Fourier_decomp} \\
$SFR$  & star formation rate & \ref{sec:parent_sample}\\ 
$\Sigma_{*,ct}$ & stellar mass surface densities averaged within the central 3 arcsec &  \ref{sec:parent_sample}\\ 
$z$ &  redshift  & \ref{sec:parent_sample}\\ 

\hline
{\bf Mathematical terms:} & &\\
K-S test   & Kolmogorov-Smirnov test & \ref{sec:subsample} \\
 $P_{KS}$ & K-S test probabilities & \ref{sec:subsample} \\
 $\rho_{cor}$ & Pearson correlation coefficient & \ref{sec:result_relations} (Fig.~\ref{fig:prop_cor})\\

\hline
{\bf Samples:} & &\\
EHC & Extremely High Central SFR, a reference analysis sample& \ref{sec:subsample}\\
HC & High Central SFR,  one of the main analysis sample &   \ref{sec:subsample} \\
LC  &  Low Central SFR, one of the main analysis sample &   \ref{sec:subsample} \\
($g-r$)-control sample &  control samples that are matched in global $g-r$ to the analysis samples & \ref{sec:control_sample} \\

\hline
{\bf References:} &  & \\
A92b & \citet{Athanassoula92b} &  \ref{sec:introduction} \\
W12  &  \citet{Wang12} &  \ref{sec:introduction} \\ 

\end{tabular}
}
\label{tab:para}
\end{table}

\end{document}